\documentclass[twocolumn]{bmcart}

\usepackage{amsthm,amsmath}
\RequirePackage[numbers]{natbib}
\RequirePackage{hyperref}
\usepackage[utf8]{inputenc} 

\usepackage[para,online,flushleft]{threeparttable}

\def\includegraphics{}

\startlocaldefs
\usepackage{siunitx}
\usepackage{xspace}
\usepackage{csquotes}
\usepackage{algorithm}
\usepackage{enumitem}
\floatname{algorithm}{Protocol}
\usepackage{algpseudocode}
\usepackage{amssymb}
\usepackage{booktabs}
\usepackage{graphics}
\usepackage{pgfplotstable}
\usepackage{cleveref}

\newcommand{\ourname}{SPIKE}
\newcommand{\suppl}{Appendix}

\newcommand{\fig}[1]{Figure~\ref{#1}} 
\newcommand{\tab}[1]{Table~\ref{#1}}
\newcommand{\tabs}[1]{Tables~\ref{#1}}
\newcommand{\prot}[1]{Protocol~\ref{#1}}
\newcommand{\sprot}[1]{Subprotocol~\ref{#1}}
\newcommand{\prots}[1]{Protocols~\ref{#1}}
\newcommand{\lnpl}[1]{Lines~\ref{#1}} 
\newcommand{\lnsg}[1]{Line~\ref{#1}}

\newcommand{\ANDGate}{$\texttt{AND}$}
\newcommand{\XORGate}{$\texttt{XOR}$}
\newcommand{\ORGate}{$\texttt{OR}$}
\newcommand{\NOTGate}{$\texttt{Not}$}
\newcommand{\MUXGate}{$\texttt{MUX}$}
\newcommand{\ADDGate}{$\texttt{ADD}$}

\newcommand{\mpc}{MPC\xspace}
\newcommand{\he}{Homomorphic Encryption\xspace}
\newcommand{\spdz}{MP-SPDZ\xspace}
\newcommand{\bss}{$\mathcal{B}$\xspace}
\newcommand{\bssform}{\mathcal{B}}
\newcommand{\ass}{$\mathcal{A}$\xspace}
\newcommand{\assform}{\mathcal{A}}
\newcommand{\gc}{$\mathcal{Y}$\xspace}
\newcommand{\gcform}{\mathcal{Y}}

\newcommand{\ce}[1]{\ensuremath\mathsf{#1}} 
\newcommand{\cf}[1]{\mathtt{#1}} 
\newcommand{\atob}{\cf{a2b}}
\newcommand{\btoa}{\cf{b2a}}

\DeclareSIUnit{\bit}{b}
\DeclareSIUnit{\byte}{B}
\newcommand{\Gbits}[1]{\SI{#1}{\giga\bit/\second}}

\newcommand{\tern}[3]{#1\textbf{ ? }#2\textbf{ : }#3}
\newcommand{\sesh}[2]{\langle #1 \rangle^{#2}}
\newcommand{\seshce}[2]{\langle\ce{#1}\rangle^{#2}}

\endlocaldefs

\begin{document}

\begin{frontmatter}

\begin{fmbox}
\dochead{Research}


\title{SPIKE: Secure and Private Investigation of the Kidney Exchange problem}


\author[
  addressref={aff1},                   
 noteref={n1},                        
  email={timm.birka@stud.tu-darmstadt.de}   
]{\inits{T.}\fnm{Timm} \snm{Birka}}
\author[
  addressref={aff2},
  email={hamacher@bio.tu-darmstadt.de}
]{\inits{T.}\fnm{Kay} \snm{Hamacher}}
\author[
  addressref={aff2},
  email={kussel@cbs.tu-darmstadt.de}
]{\inits{T.}\fnm{Tobias} \snm{Kussel}}
\author[
  addressref={aff1},corref={aff1},
  email={moellering@encrypto.cs.tu-darmstadt.de}
]{\inits{T.}\fnm{Helen} \snm{Möllering}}
\author[
  addressref={aff1},
  email={schneider@encrypto.cs.tu-darmstadt.de}
]{\inits{T.}\fnm{Thomas} \snm{Schneider}}


\address[id=aff1]{
  \orgdiv{ENCRYPTO},             
  \orgname{Technical University of Darmstadt},          
  \city{Darmstadt},                              
  \cny{Germany}                                    
}
\address[id=aff2]{%
  \orgdiv{Computational Biology \& Simulation group},
  \orgname{Technical University of Darmstadt},          
  \city{Darmstadt},                              
  \cny{Germany}   
}


\begin{artnotes}
\note[id=n1]{Lead Author} 
\end{artnotes}



\begin{abstractbox}

\begin{abstract}
\parttitle{Background}
The kidney exchange problem (KEP) addresses the matching of patients in need for a replacement organ with compatible living donors. Ideally many medical institutions should participate in a matching program to increase the chance for successful matches. However, to fulfill legal requirements current systems use complicated policy-based data protection mechanisms that effectively exclude smaller medical facilities to participate. Employing secure multi-party computation (\mpc) techniques provides a technical way to satisfy data protection requirements for highly sensitive personal health information while simultaneously reducing the regulatory~burdens. 
\parttitle{Results}
We have designed, implemented, and benchmarked \ourname, a secure \mpc-based privacy-preserving KEP which computes a solution by finding matching donor--recipient pairs in a graph structure. \ourname{} matches 40 pairs in cycles of length 2 in less than 4 minutes and outperforms the previous state-of-the-art protocol by a factor of $400\times$ in runtime while providing medically more robust solutions. 
\parttitle{Conclusions}
We show how to solve the KEP in a robust and privacy-preserving manner achieving practical performance. The usage of \mpc techniques fulfills many data protection requirements on a technical level, allowing smaller health care providers to directly participate in a kidney exchange with reduced legal processes.
\end{abstract}




\end{abstractbox}
\end{fmbox}

\end{frontmatter}



\section*{Introduction}
Around \SI{7}{\percent} of U.S. adults are affected by chronic kidney
disease~\cite{murphyTrendsPrevalence2016}. With the increasing age of the
population in most countries, end-stage renal disease constitutes a rapidly
increasing challenge for health care
systems~\cite{thurlowGlobalEpidemiology2021}. Humans are able to live a normal
life with at least one functioning kidney~\cite{ibrafoltan09}.  However, when
both kidneys of a person are malfunctioning, this person requires kidney
replacement therapy to survive, i.e., either dialysis or the donation of a
functioning kidney.

Transplantations of deceased donor organs unfortunately imply long waiting times
as transplant waiting lists grow, given that the number of donations
significantly exceed supply~\cite{euro2021Annual}. The other option is to find a
living person that is willing to donate one of  their kidneys. In general,
living donor donations result in shorter waiting times and tend to have better
long term outcomes compared to deceased donor donations~\cite{nein14}.
Unfortunately, finding a willing, living donor does not guarantee (medical)
compatibility with the recipient. Hence, the living donor exchange system was
introduced in 1991~\cite{ellisonSystematicReview2014}, which allows recipients
with incompatible living donors, in the following referenced as \emph{pairs}, to
exchange their donors such that ideally each recipient can receive a compatible
kidney donation. 

In this work, we consider a scenario in which several pairs exchange their
donors in a cyclic fashion, so that each donating pair receives a compatible
kidney. These cycles are called \emph{exchange cycles}~\cite{biro21}.

\begin{figure}
    \includegraphics[width=.468\textwidth]{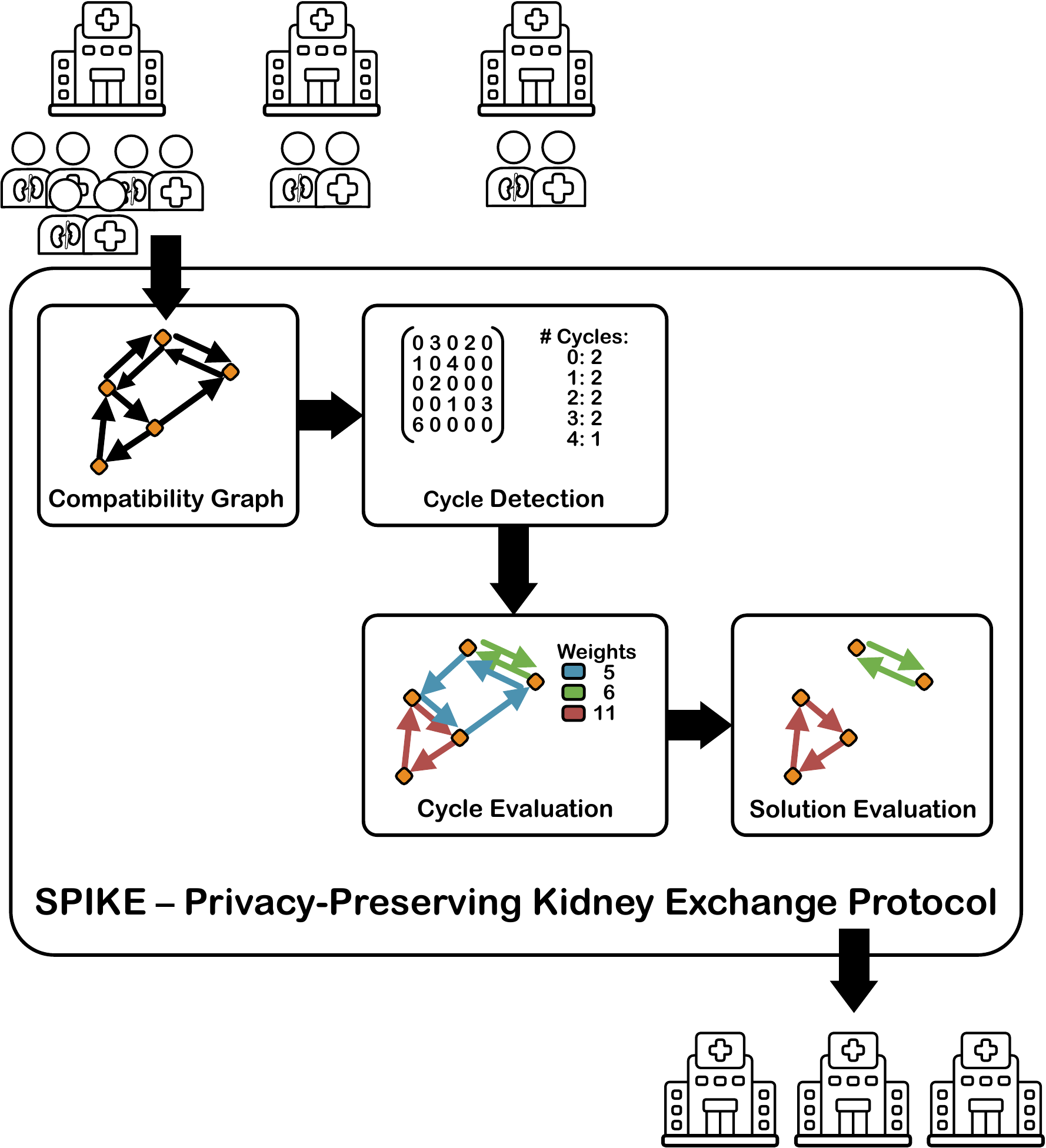}
    \caption{Overview of our Privacy-Preserving Kidney Exchange Protocol SPIKE.
    The best set of exchange cycles are calculated, while the patients' data
  remain strictly private.}%
    \label{fig:highlevel}
\end{figure}

As a first step for finding possible exchange cycles, we have to evaluate the
donors' and recipients' medical data to determine compatibility between pairs.
Afterwards, we have to identify possible exchange cycles. This problem is known
as the kidney exchange problem (KEP)~\cite{biro21} and can be described as
finding cycles in a directed graph where each vertex represents a pair and a
directed edge describes the compatibility between two pairs. A schematic view of
the protocol can be seen in \fig{fig:highlevel}.

The process requires the analysis of highly sensitive medical health data, which
makes it crucial that no information is leaked accidentally or to unauthorized
personnel. Thus, KEP requires the implementation of strong privacy-preserving
solutions where the plaintext health information remains locally at each medical
institution and the analysis is only run on \enquote{encrypted} data which is
leaking nothing beyond the output: an exchange cycle with high transplantation
success likelihood.\footnote{This cycle still requires a final check by medical
experts.} Note that such a distributed solution also enhances security against
data breaches as having to attack multiple parties is
significantly harder than a single target. Similarly, it also simplifies the
compliance with regulatory requirements potentially complicating or even
prohibiting data sharing among facilities.

\subsection*{Contributions and Outline} 
In this work, we provide the following contributions: 
\begin{itemize}
    \item \textbf{Efficient Privacy-Preserving Kidney Exchange protocol:} We
      design and implement \ourname, an efficient, distributed,
      privacy-preserv-ing protocol for solving the kidney exchange problem in the
      semi-honest security model. In contrast to the current
      state-of-the-art~\cite{breuer20,breuer2022}, \ourname{} improves the
      medical compatibility matching by considering additional factors, namely,
      age, sex, human leukocyte antigens, and weight, that significantly affect
      compatibility between potential donors and recipients and is, thus, more
      robust than previous solutions by reducing the risk of failing procedures.
    \item \textbf{Comprehensive Empirical Evaluation:} We implement and
      extensively benchmark \ourname{} and show that it has practical runtimes
      and communication costs. We achieve about $\num{30000}\times$ speedup
      over~\cite{breuer20} and $\num{400}\times$ over~\cite{breuer2022} thanks
      to our carefully optimized hybrid secure multi-party computation (\mpc) protocols. Further, we
      provide additional \mbox{(micro-)} benchmarks and network settings to
      further demonstrate scalability and practicality of \ourname.
    \item \textbf{Open-source Implementation:} \ourname{} is available under the GNU LGPL v3 license\footnote{\url{https://www.gnu.org/licenses/lgpl-3.0}} here: \url{https://encrypto.de/code/PPKE}.
\end{itemize}

\section*{Related Work}
In this section, we summarize the related work on the Kidney Exchange Problem
(KEP) with and without considering data privacy.

\subsubsection*{Robust KEP}
One major issue in kidney exchange programs is the potential cancellation of
transplantations \emph{after} having already determined exchange cycles of
compatible donors and recipients. Reasons for such cancellations are manifold,
e.g., a donor withdraws his consent as his non-compatible relative has already
received a kidney via the waiting list from a deceased donor in the
meantime~\cite{pansart2014}. These issues call for \emph{robust} solutions to
the KEP, i.e., flexibility for recipient/donor dropouts and including as much as
possible medical factors that can be algorithmically evaluated.

Carvalho et al.~\cite{carvalho2020} propose three policies that are able to cope
with dropouts within an kidney exchange cycle. One takes the costs (or missed
gains) of planned transplants that do not proceed into account to find a
solution with high probability of being successfully executed. The other two
policies investigate strategies for recovering exchange cycles after dropouts. The plaintext algorithms in~\cite{carvalho2020} are computationally expensive
and, thus, cannot be trivially realized as secure computation protocols. 

Ashby et al.~\cite{ashby2018} introduce a calculator for determining
compatibility in kidney exchange which they use to evaluate the importance of
various medical factors such as age, sex, obesity, weight, height, human leukocyte antigen (HLA)
mismatches and ABO blood groups (see Section \enquote{Medical Background}). In our work, we increase the robustness of our privacy-preserving kidney exchange protocol by
including the additional important biomedical factors from~\cite{ashby2018}.
Furthermore, we recommend to use cycle sizes of two or three to reduce the
impact of withdrawals~\cite{pansart2014}. The size is also beneficial for
practical considerations with respect to medical staff and other resources
needed for transplantations as all operations of one exchange cycle should
ideally be executed simultaneously. This recommendation reflects current best
practices~\cite{abraham2007}.

\subsubsection*{Privacy-preserving KEP}
Just two works, both by Breuer et al.~\cite{breuer20,breuer2022}, investigate
how to solve the kidney exchange problem in a decentralized privacy-preserving
manner. Both consider the semi-honest security model.

\textit{Privacy-preserving KEP with HE.} The first protocol~\cite{breuer20} uses
homomorphic encryption (concretely, a threshold variant of the Paillier
cryptosystem~\cite{fouque2000}). It instantiates a computing party for each pair
of a non-compatible donor and recipient at the providing hospital, thus,
effectively creating a multi-party computation (\mpc) protocol.

The protocol first pre-computes a set of all possible exchange constellations
independent of any input data. Cycles of all lengths up to $3$ are computed (but an
arbitrary value could be chosen). Next, the pairs jointly compute an adjacency
matrix with the pair-wise compatibility based on HLA crossmatching and ABO blood
groups. Combining the results with the exchange constellations, the graph with
the maximal size is delivered as the output. The protocol's runtime scales
exponentially with the number of pairs: starting with a runtime of $14$ seconds
for two pairs it increases to $13$ hours for nine pairs. Unfortunately, such
runtimes are prohibitive for practical deployment.

\textit{Privacy-preserving KEP with Shamir's Secret Sharing.} In a concurrent
work to ours, Breuer et al.~\cite{breuer2022} introduced a privacy-preserving
KEP protocol for crossover kidney exchanges with polynomial computation
complexity. \enquote{Crossover} hereby means that the kidney exchange is done
among two pairs, i.e., the exchange cycle size is limited to two in contrast
to~\cite{breuer20}. This limitation, however, enables a significant efficiency
improvement for matchings with more than 13 pairs. For example, with 15 pairs it
reduces the runtime of the old protocol~\cite{breuer20} from 8.5 hours to 30 minutes. Additionally, the
new protocol enables a dynamic setting where donor-recipient pairs can be
added/removed from the exchange graph at any point in time. On the technical
side, the authors replace HE and fully rely on a \mpc-technique called Shamir's
Secret Sharing (SSS)~\cite{shamir1979share} implemented with the MP-SPDZ
framework~\cite{keller2020}. Beyond the dynamic setting and the change to \mpc,
the new protocol employs the graph matching algorithm by Pape and
Conradt~\cite{pape1980} for better efficiency in the matching between compatible
donors and recipients. 

Our privacy-preserving KEP protocol \ourname{} offers practical runtimes for
real-world deployment. Our runtimes significantly outperform the measured runtimes of previous works~\cite{breuer20,breuer2022}, e.g., by a factor of
hundreds/thousands for 9 recipient-donor pairs with a cycle length of 2. This is due to an efficient symbiosis of three \mpc
techniques and protocol optimizations that we will detail in the next section.
Furthermore, we improve the robustness of \ourname{} by including four
additional biological factors notably impacting the transplantation success
rate~\cite{ashby2018}. Thus, our protocol focuses on high medical quality rather
than pure size while also significantly improving efficiency.

\section*{Background}%
\label{sec:background}

In this work, we present a privacy-preserving solution to the
\emph{kidney exchange problem} (KEP). We interpret the KEP as an optimization
problem, specifically finding cycles with a maximal coverage of nodes on a
compatibility graph and a maximal aggregated edge weight. The graph is
constructed according to medical compatibility factors. This section gives the
required background information to understand the underlying aspects of
biomedicine, graph theory, as well as the used privacy-preservation techniques
of secure multi-party computation (\mpc).

\subsection*{Medical Background} \label{medical}
In the following, we introduce the medical background, i.e., biological factors
used in our protocol that cause general immunological incompatibility or
influence success likelihood for a kidney transplantation.

\subsubsection*{General immunological compatibility}

While many medical factors are involved in the definite assessment of
donor-recipient compatibility, some can be algorithmically determined.
For example, one key factor in avoiding allograft rejection---immunological
compatibility---can be evaluated following evidence-based guidelines. Our kidney
exchange protocol uses a specific form of immunological compatibility, the HLA
crossmatch, as a transplant prohibiting factor.

\label{hla_prelims}
\noindent\textit{Human Leukocyte Antigens crossmatch.} The human immune
system is responsible for the protection of the organism against potentially
harmful invaders (called \emph{pathogens}). Antigens are molecular structures often found on
the surface of pathogens, but also naturally occurring in the body. \emph{Antibodies} can attach to those structures,
preventing the pathogens from docking, thus inhibiting their harmful effect. One
important group of endogenous antigens which occur in varying numbers in every human,
forming the immunological \enquote{fingerprint} the immune system recognizes as
normal, are the \emph{human leukocyte antigens}. Out of the three classes
of HLA~\cite{sung2007}, only classes I and II are of interest in this work. 

With a \emph{HLA crossmatch} general compatibility between recipient and donor
can be determined: The human leukocyte \emph{antigens} of a donor are matched
against existing human leukocyte \emph{antibodies} of a possible
recipient~\cite{euro102021}. HLA crossmatch positive kidney transplants carry a
significantly higher risk of antibody-mediated rejection or allograft rejection
due to already existing antibodies~\cite{lefaucher2010, ntokou2011}. Modern
immunosupressants might make such a procedure possible~\cite{santos2014}, but
those cases require specialized, in-depth medical assessment and are out of
scope of a general, algorithmic evaluation.

Following Eurotransplant's guidelines~\cite{euro102021}, we consider HLA groups,
which are also most frequently screened in preparation for kidney replacement
therapy~\cite{euro42021}: the HLA encoded at HLA-A, -B, and -DR loci.
Additionally, we consider the HLA-DQ coded antigens, which are related to some
cases of antibody-mediated rejection~\cite{leeaphorn18}. The full list of HLA
considered in \ourname{} can be seen in \tab{tab:hla_classes}.
\begin{table}[H]
  \setlength\extrarowheight{2pt}
\caption{HLA split antigens assessed for biomedical donor -- recipient
compatibility testing in \ourname.}%
\label{tab:hla_classes}
\begin{tabular}{lllll}\toprule
\multicolumn{3}{c}{Class I}                           & \multicolumn{2}{c}{Class II}\\ \cmidrule(r){1-3} \cmidrule(l){4-5} 
\multicolumn{1}{c}{HLA-A} & \multicolumn{2}{c}{HLA-B} & \multicolumn{1}{c}{HLA-DR} & \multicolumn{1}{c}{HLA-DQ} \\ \midrule
A23                       & B38         & B60         & DR11                       & DQ5                        \\
A24                       & B39         & B61         & DR12                       & DQ6                        \\
A25                       & B44         & B62         & DR13                       & DQ7                        \\
A26                       & B45         & B63         & DR14                       & DQ8                        \\
A29                       & B49         & B64         & DR15                       & DQ9                        \\
A31                       & B50         & B65         & DR16                       &                            \\
A32                       & B51         & B71         & DR17                       &                            \\
A33                       & B52         & B72         & DR18                       &                            \\
A34                       & B54         & B75         &                            &                            \\
A66                       & B55         & B76         &                            &                            \\
A68                       & B56         & B77         &                            &                            \\
A69                       & B57         &             &                            &                            \\
A74                       & B58         &             & &   \\\bottomrule
\end{tabular}
\end{table}

\subsubsection*{Match quality estimation}
Additionally to the previously introduced procedure that prevents immunological
incompatibility, we strive to find the medically best/most robust solution to
the kidney exchange problem -- that includes maximal survival probability. For
that, we calculate a match quality index, based on the following additional medical~factors:

\begin{enumerate}[label=(\roman*)]
    \item \noindent\textit{HLA match.}

Additionally to the HLA crossmatch, HLAs influence the probability of a
successful transplantation. Concretely, it increases if the donor has a subset
or the same HLA as the recipient. The number of \enquote{mismatches} is
associated with increased allograft rejection rates, as the probability that a
recipient develops antibodies to those mismatched antigens
increases~\cite{opelz12}. HLA mismatches do not constitute exclusion criteria,
as immunosupressants can reduce the rejection probability. The use of
immunosupressants, however, is itself linked to higher rejection
rates~\cite{opelz1997,opelz12,lim12}. Special importance comes to the HLA-DQ
group, as mismatches of it are strongly linked to antibody-mediated
rejections~\cite{leeaphorn18}.

Each person can inherit up to two types of HLA per group. Hence, at most two
mismatches can occur per group~\cite{nguyen2021}. The impact of HLA mismatches
can be categorized in four bins: having no mismatch, a very rare case and mostly
occurring in twin donor-recipient pairs, having \numrange{1}{2} mismatches,
having \numrange{3}{4} mismatches, and, worst of all, having more than \num{5}
mismatches~\cite{opelz12}. The last group shows a more than \SI{6}{\percent}
cumulative risk for death with a functioning graft during the first year. We
weight HLA mismatches according to those four categories.

\item \noindent\textit{ABO blood type.}\label{abo_prelims}
The ABO blood type system is based on the presence or absence of two types of
antigens on the surface of the red blood cells~\cite{blutspenden21}. The absence
of both type A and type B antigens mark blood type O, the presence of both mark
blood type AB, and the presence of only one mark blood type A and B,
respectively. Receiving blood with an incompatible blood type leads to blood
clumping due to an immune reaction and a possibly failed procedure. Compatible
pairings are given in \tab{tab:abo_comp}.

By pre-processing the donor organ, grafts from ABO incompatible donors are
possible~\cite{weerd18}, although linked to severe adversary reactions during
the first year post transplantation. These reactions include a higher risk of
allograft loss, severe viral infections, antibody-mediated rejections, and
postoperative bleeding. After this first year, however, the long-term survival
rate is comparable to ABO compatible transplants~\cite{weerd18}.

\begin{table}[H]
        \caption{ABO compatibility~\cite{blutspenden21}}\label{tab:abo_comp}
        \begin{tabular}{ccc}
            \toprule
            Blood Group & Can Receive From & Can Donate To \\
            \midrule
            O & O & O, A, B, AB \\
            A & O, A & A, AB \\
            B & O, B & B, AB \\
            AB & O, A, B, AB & AB \\
            \midrule
        \end{tabular}
        \label{tab:abo}
\end{table}

\item \noindent\textit{Age.}\label{age_prelims}
According to Waiser et al.~\cite{waiser00}, also age disparity influences
allograft survival post transplant. The authors examined the role of age of the
donor and recipient using two categories: \emph{junior} participants aged below
\num{55} years and \emph{seniors} participants older than \num{55} years. The
results show that intra-categorical transplants show the best outcomes, followed
by pairings of junior donors and senior recipients. The worst outcomes were
observed for pairings with senior donors and junior recipients.

\item \noindent\textit{Sex.}\label{sex_prelims}
As shown by Zhou et al.~\cite{zhoua13}, the combination of donor-recipient
sexes impact the transplant success probability. The worst allograft survival
rates were observed in male recipients for female donor organs, while same-sex
pairs performed slightly better than female recipients for male donor organs.

\item \noindent\textit{Weight.}\label{size_prelims}
Recipients who received a kidney from a donor who weighs less have higher
chances of allograft loss than other recipients~\cite{miller17}.
El-Agroudy et al.~\cite{elagroudy03} reason that the
allograft loss for recipients with kidneys from lighter donors might be caused
by the kidney being unable to support the body functions of a heavier recipient.
\end{enumerate}

\subsection*{Graph Theory}
We represent the structure of the kidney exchange problem (KEP) as a
(bipartite) graph problem. A graph $\mathcal{G}$ consists of a set of vertices
$\mathcal{V}$ and an edge set~$\mathcal{E}$ connecting the vertices.
Technically, we deal with a \emph{bipartite} graph, i.e., consisting of two
different sets of vertices (donors and recipients), but as those register
pairwise for the kidney exchange, we can \enquote{collapse} each
donor-recipient pair into one vertex in $\mathcal{V}$. If two vertices $v, u
\in \mathcal{V}$ are connected by an edge, then $(v, u) \in \mathcal{E}$. 
We consider a directed graph with directed edges from $v$ to $u$.
Furthermore, we use \emph{weighted} edges by associating a
scalar weight to each edge, according to its \enquote{importance} in the
network. The weights represent the degree of medical compatibility. We only
allow positive edge weights.

Our goal is to find all cycles within the graph. A cycle $c$ is a list of
vertices $\{v_1, v_{2}, ..., v_m\}$ where an edge exists from vertex~$v_i$ to~$v_{i+1}$ for $i \in \{1, ..., m-1\}$ and, to close the \enquote{loop}, from
vertex $v_m$ back to vertex~$v_1$. In a vertex disjoint cycle, each vertex
appears at most once within the cycle. We define the length of a cycle as the
number of edges that are used to form that cycle.

One representation of a (weighted) graph structure is the
\emph{adjacency matrix}, a square matrix $A$ with one row/column for each
vertex. If an edge exists between vertices $i$ and $j$, then, the entry
$a_{ij}=w$, with $w>0$ being the edge weight and $a_{ij}=0$ otherwise.
This work uses the fact that by raising the adjacency matrix to the $\ell$th
power, one can quickly compute the number of paths between two vertices with a
given length $\ell$. That means, that vertices $i$ and $j$ are connected by
$(A^\ell)_{ij}$ paths of length $\ell$. The diagonal elements give the number of
cycles of length $\ell$ by finding paths starting and ending on the same vertex.

\subsection*{Secure Computation}
\label{sub:secure_computation}
Secure computation techniques enable multiple parties to securely evaluate an
arbitrary function on their private inputs. Ideally nothing is leaked beyond what
can be inferred from the output. A secure computation protocol must be able to
realize this functionality without relying on a trusted party. To verify its
security, it is typically compared to the so-called \emph{ideal functionality}
which is a trusted third party that runs the computation on behalf of the data
owners.

Privacy research has mainly worked on two paradigms for secure computation:
Homomorphic Encryption~(HE) and Secure Multi-Party Computation (\mpc). HE
schemes are special public-key encryption schemes that allow to realize
(some limited) mathematical operations under encryption. However, they
tend to be computing intensive making them (yet) often unsuitable for
real-world applications. In contrast, \mpc techniques are typically more
efficient with respect to computation as they are mainly based on efficient symmetric
encryption. Additionally, \mpc protocols can compute arbitrary functions. They
are typically split into a setup and an online phase where the setup phase is
independent of the input data and, thus, can be pre-computed. This separation
enables to significantly speed up the time-critical online phase as
pre-computation can be done in idle times when input data is not yet available.
However, \mpc involves two or more parties who jointly evaluate the desired
function in a secure manner, hence, it requires communication among the parties.
Both paradigms have already been used in the context of privacy-preserving
genome-wide association studies~\cite{cho2018secure,bonte2018towards,tkachenkoLargeScalePrivacyPreserving2018a} as
well as other applications in the health care area~\cite{ST19,gunther2020pem,BFKLSS09,BFLSS11}.

To have provably secure privacy guarantees while achieving practical efficiency,
\ourname{} efficiently combines multiple \mpc techniques which we introduce in
the following.

\subsubsection*{Secure Multi-Party Computation (\mpc)}
Introduced by Andrew Yao's seminarial work \enquote{How to Generate and Exchange
Secrets}~\cite{yaoHowGenerate1986} in 1986, secure Multi-Party Computation
(\mpc) was considered a theoretical field first. \mpc are cryptographic
protocols that can securely compute an arbitrary function among two or more
parties on their private inputs. Enabled by the rapid development of
computer hardware and the development of the first \mpc compiler
\enquote{Fairplay}~\cite{malkhiFairplaySecure2004}, first practical uses were
demonstrated around the year 2004. Since then, \mpc is a flourishing research
field and due to novel protocols and optimizations, such as \enquote{Free
\XORGate}~\cite{kolesnikovImprovedGarbled2008} or
\enquote{Halfgates}~\cite{zahurTwoHalves2015}, practical applications in many
fields were shown~\cite{ST19,patra2021aby2,jarvinen19}. 

In this work, we rely on three well established secure \emph{two-party}
computation techniques, i.e., the secure computation protocols are run among
exactly two parties: Arithmetic Secret Sharing (\ass), Boolean Sharing (\bss),
both based on~\cite{goldreichHowPlay1987}, and Yao's Garbled Circuits~(Y),
originally introduced in~\cite{yaoHowGenerate1986}. Each technique differs in
how it creates (\emph{shares}) and reconstructs secrets, but also how
(efficiently) certain types of operations can be realized.

\noindent\textit{Notation.} In the following, $\langle x \rangle_{i}^s$
denotes a secret share of $x$ shared using \mpc technique $s\in\{A,B,Y\}$ and
held by party $P_i$, where $i\in\{0,1\}$. 

\noindent\textit{Yao's Garbled Circuits (\gc).}\label{para:yaosGC} Yao's Garbled
Circuits enable two parties, called the \emph{garbler} and the \emph{evaluator},
to securely evaluate a function $f$ represented as \emph{Boolean circuit}, i.e.,
a directed acyclic graph where the nodes are logic gates and the edges (called
\emph{wires}) are the Boolean in- and outputs. For functional completeness
\ANDGate{} and \XORGate{} gates are sufficient. The garbler generates random
keys for each possible state of each wire $k^w_0,k^w_1\in\{0,1\}^\kappa$, where
$\kappa$ is the symmetric security parameter (set to $\kappa=128$ in our implementation) and $w$ is the respective wire. For all input combinations of each gate in the circuit, it
uses the input keys to encrypt the corresponding output key
(cf.~\tab{fig:ygc}). The order of the four ciphertexts is then permuted randomly and the \emph{garbled circuit} is sent to the evaluator together with the keys associated to the garbler's
input. As those keys look random, the evaluator cannot extract any information about
the input of the garbler. Next, the evaluator engages in an oblivious
transfer~\cite{wiesnerConjugateCoding1983,rabinHowExchange1981} to receive the
keys for his input without revealing it to the garbler. Equipped with all keys, it evaluates
the garbled circuit to receive the circuit's output keys which the parties
jointly decode. Thanks to several optimizations,
e.g.,~\cite{kolesnikovImprovedGarbled2008, zahurTwoHalves2015,
asharovMoreEfficient2017}, \gc requires no communication for the evaluation of
an \XORGate{} gate and only $1.5\kappa$ bits of communication for \ANDGate{}
gates. \gc needs a constant number of communication rounds independent of the
circuit depth.

\begin{table}
  \caption{Garbled \ANDGate{} Gate}%
  \label{fig:ygc}
  \begin{tabular}[t]{cccc} \toprule
  Input $w_0$&Input $w_1$&Output $w_2$&Garbled Value\\ \cmidrule(r){1-2}\cmidrule{3-3}\cmidrule(l){4-4}
  $k_0^{w_0}$&$k_0^{w_1}$&$k_0^{w_2}$& $Enc_{k_0^{w_0},k_0^{w_1}}(k_0^{w_2})$\\
  $k_0^{w_0}$&$k_1^{w_1}$&$k_0^{w_2}$& $Enc_{k_0^{w_0},k_1^{w_1}}(k_0^{w_2})$\\
  $k_1^{w_0}$ &$k_0^{w_1}$&$k_0^{w_2}$& $Enc_{k_1^{w_0},k_0^{w_1}}(k_0^{w_2})$\\
  $k_1^{w_0}$ &$k_1^{w_1}$&$k_1^{w_2}$& $Enc_{k_1^{w_0},k_1^{w_1}}(k_1^{w_2})$\\
  \bottomrule
\end{tabular} \end{table}%

\noindent\textit{Boolean and Arithmetic Secret Sharing
(\bss/\ass).}\label{para:gmw} In Additive Arithmetic Secret Sharing
(\ass) operations on $\ell$-bit inputs are done in an algebraic ring $\mathbb{Z}_{2^{\ell}}$,  where $\ell$ is the bit length. Although the
technique can also be used among an arbitrary number of
parties~\cite{benor1988}, we focus here on the two party setting as introduced by
Goldreich et al.~\cite{goldreichHowPlay1987}.

To share a secret value $x$, party $P_i$, $i\in\{0,1\}$, generates a
random value $r\in_R \mathbb{Z}_{p}$ and sets its arithmetic share to $\langle
x\rangle^A_{i}=r$. Then, $P_i$ also determines party $P_{1-i}$'s share
$\langle x\rangle^A_{1-i}=x-r \mod 2^{\ell}$ and sends it to $P_{1-i}$. To
reconstruct the secret, one needs to know \emph{both} shares and compute
$x=\langle x\rangle^A_0+\langle x\rangle^A_1 \mod 2^{\ell}$. Boolean Secret
Sharing (\bss) describes the special case where $\ell=1$, viz.
$\mathbb{Z}_2=\left\{0,1\right\}$.

Note that a share $\langle x\rangle^A_i$ (resp. $\langle x\rangle^B_i$) is
random and does not leak anything about the secret $x$. Secure addition
(respectively, \XORGate ing in \bss) can be executed locally, that is without
communication between the parties. Secure multiplication (respectively,
\ANDGate{} in \bss) is done in an interactive protocol among the two parties
using so-called multiplication
triples~\cite{beaver1991efficient,dsz15,rathee2019improved}. Using only addition
and multiplication (similarly, \ANDGate{} and \XORGate) arbitrary functions can
be calculated.

\noindent\textit{ABY Framework. }\label{para:aby} All three \mpc
techniques are implemented in the state-of-the-art secure two-party computation
framework ABY~\cite{dsz15} which we use in our experiments. Additionally, ABY
also contains efficient conversions between them and supports Single Instruction
Multiple Data (SIMD) operations to parallelize identical operations on different
data while reducing memory usage and~runtime. Arithmetic Secret Sharing in ABY
is performed on the ring $\mathbb{Z}_{2^\ell}$, that is with $2^\ell$ elements
where $\ell$ is the bitlength of the data type (most often $\ell =
\SI{32}{bit}$). A recent work by Patra et al.~\cite{patra2021aby2}
improves~\cite{dsz15} by making the online communication of scalar multiplication independent of the vector
dimensions and reducing online communication for \ANDGate{} gates with two
inputs in \bss{} by a factor of 2. Unfortunately, these protocols have been implemented only very recently in MOTION2NX~\cite{BCS21PriMLNeurIPS} which is why we use~\cite{dsz15} in our implementation.

\subsubsection*{Security Model} In our work, we consider the
\emph{semi-honest} security model where all involved parties are assumed to be honestly following the
protocol while trying to learn as much information as possible. By
\enquote{honestly following the protocol} we, thereby, mean that they adhere to
the specifications of the protocol, e.g., they do not manipulate local
calculations or provide inconsistent data. Additionally, the parties are assumed
to not collude. This security model provides protection against curious
personnel or accidental data leakage. Although weaker than the malicious
security model where the parties might arbitrarily deviate from the protocol,
the semi-honest security model is sufficient for our use case, as hospitals are
generally trusted but legally not allowed to simply share the data among each
other. Furthermore, the semi-honest security model enables significantly more
efficient computation than the malicious model and, hence, provides a good
efficiency-privacy trade-off. Previous works on privacy-preserving KEP~\cite{breuer20,breuer2022}
are also in the semi-honest security model.

\subsubsection*{Outsourcing Data-Model}
Although we use 2-party \mpc to perform the computation, any number of parties
can provide input data. This \emph{outsourcing}~\cite{kamara2011secure} works by all data owners sharing
their data and sending one share to each of the two non-colluding computation
servers. Those servers, then, perform the actual secure computation on behalf of
the data owners while not being able to learn anything about the private input
data. This scenario has three main benefits:
\begin{itemize}
  \item The communication of $N$-party \mpc scales at least linearly, if not
    quadratic in the number of participating
    parties~\cite{damgardMultipartyComputation2012}. By outsourcing the
    $N$-party computation to $M$ parties, here $M=2$, the communication is
    improved by around $\mathcal{O}(\tfrac{N}{M})$.
  \item As the input owners do not participate in the computation itself, the
    outsourced protocol provides security against malicious data
    owners~\cite{kamara2011secure}. At most they can corrupt the correctness of
    the calculation, but not the privacy.
  \item The location of the computation servers can be chosen pragmatically,
    e.g., the two locations with the highest bandwidth or lowest latency network
    connection. Of course, the non-collusion assumption must be considered.
\end{itemize}

\section*{Methods}%
\label{sec:methods}
In this section, we first define the privacy-preserving Kidney Exchange
problem (KEP) and its requirements. Then, we present our solution, which we name
\ourname, consisting of tailored modular secure \mpc protocols and include a
complexity analysis.

\subsection*{Problem Statement}
\begin{figure}
\centering
  \includegraphics[width=0.95\columnwidth]{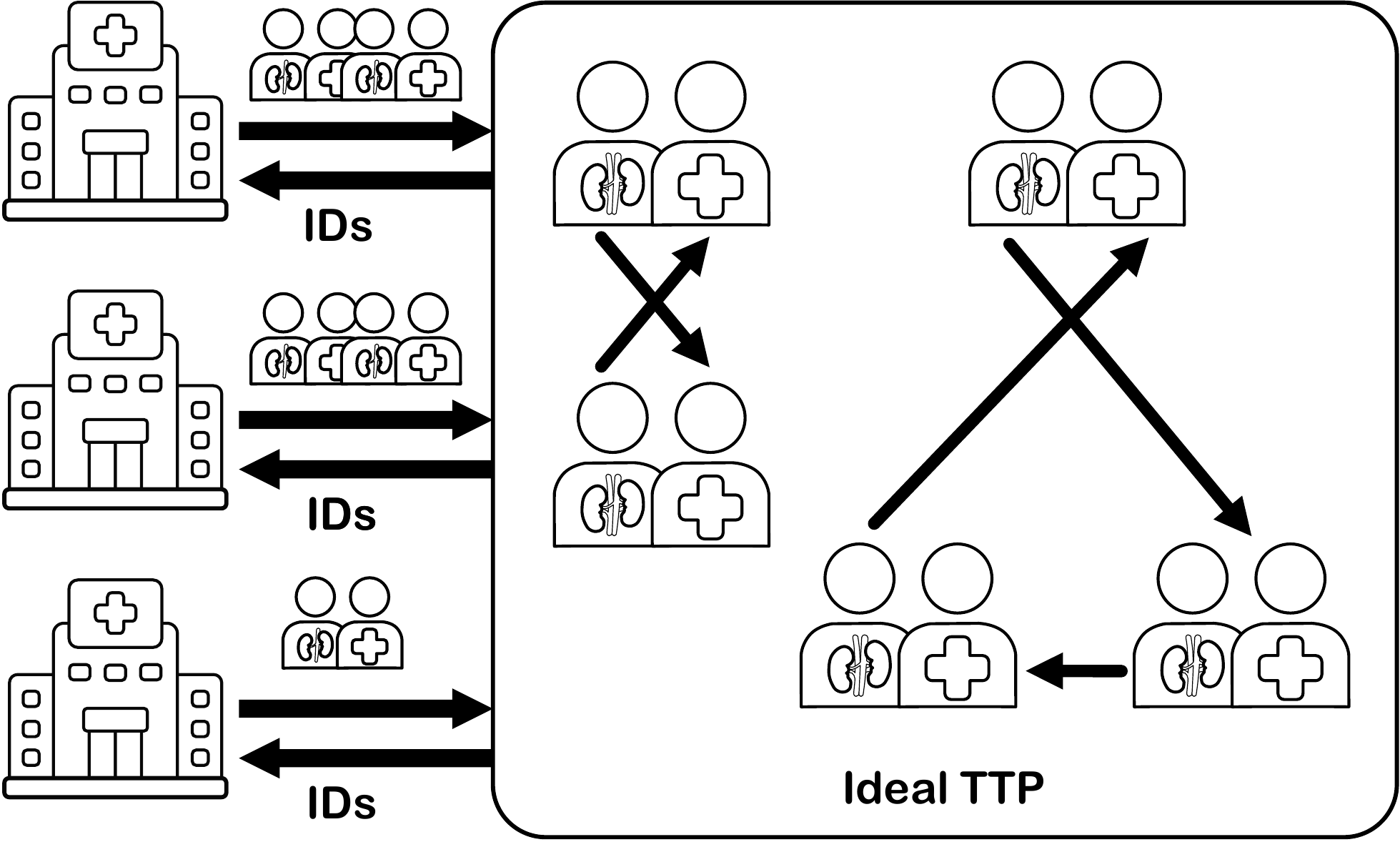}
  \caption{Ideal Functionality for a secure privacy-preserving Kidney Exchange Problem (KEP).}%
  \label{fig:idealf}
\end{figure}

\fig{fig:idealf} shows the ideal functionality for solving the
privacy-preserving KEP in a provably secure way. Assuming the (not realistic)
availability of a trusted third party (TTP), hospitals send the data of
recipients and donors to the TTP which calculates cycles of pairs of recipients
and donors with the highest probability to be compatible. Then, the TTP returns for
each recipient the information about his/her donor to the respective hospital.
Note that a final evaluation must still be done by medical experts. A
privacy-preserving KEP is meant for automatizing and, thus, accelerating, the
process of the creation of the kidney exchange cycles.

\noindent\textbf{Requirements:} We define the following requirements for a
secure privacy-preserving KEP protocol:
\begin{itemize}
    \item \textbf{Privacy.} The privacy-preserving KEP protocol must realize the
      same functionality as described in the ideal functionality while removing
      the problematic assumption of a TTP, i.e., it must leak nothing beyond
      what can be inferred from the~output.
    \item \textbf{Efficiency.} The privacy-preserving KEP protocol must be
      efficient in terms of communication and computation such that it can be
      run in reasonable time
      on standard server hardware.
    \item \textbf{Decentralization.} The privacy-preserving KEP protocol must be
      decentralized, i.e., the highly sensitive medical information of donors
      and patients must remain locally at the respective medial institution
      (inherently being compliant with the data minimisation~principle). 
    \item \textbf{Adaptability for Medical Experts.} The priva-cy-preserving KEP
      protocol must be flexibly adaptable for medical experts with respect to
      the selection of biological factors for the algorithmic evaluation of
      compatibility. They must be able to adjust the weighting between the
      included factors and cycle lengths according to state-of-the-art medical advancements. The
      protocol must be easily extendable to new factors and additional HLA
      groups.
\end{itemize}

\subsection*{\ourname: A \mpc-based privacy-preserving KEP protocol}
In this section, we provide the building blocks for our Secure and Private
Investigation of the Kidney Exchange problem: \ourname. It fulfills above requirements (see also the overview of the phases in \fig{fig:highlevel}).

First, we explain the matching phase, which analyzes the compatibility between
donors and recipients using six biological factors presented in the Background
section. Then, we continue with the determination of the number of potential
exchange cycles given a cycle length. The third phase computes the probability
of a successful transplantation based on the matching results for all potential
exchange cycles. In the final phase, we output a robust set of \emph{disjoint}
exchange cycles, i.e., with a high probability for compatibility. By design,
\ourname{} enables usage in a dynamic setting similar to the protocol
in~\cite{breuer2022}.

\noindent \emph{Notation.} 
We use Boolean operators to concisely present our \mpc protocols: $\land$ is
\ANDGate, $\lor$ is \ORGate, $\lnot$ is \NOTGate, and $\oplus$ is \XORGate. 0/1
are False/True. $|x|$ indicates the length of a vector $x$, i.e., the number of elements.
Non trivial variable names in protocols are written in
$\ce{sans~serif}$, function names (and calls) $\cf{monospaced}$.
Branching, implemented with \MUXGate{} (multiplexer) Gates, is displayed using 
ternary notation: condition \textbf{?} true statement \textbf{:} false statement.

\subsubsection*{Compatibility Matching}\label{sub:matching}
The first phase of \ourname{} is called \emph{compatibility matching}. In this
phase, we compare the pair-wise general compatibility and match quality of all
donors and recipients with respect to human leukocyte antibodies and antigens,
ABO blood group compatibility, age, sex, and weight. The output of this phase is
a weighted compatibility graph where the edge weights indicate the probability
of compatibility for each pair.

\floatname{algorithm}{Subprotocol}
\begin{algorithm}
    \caption{$\cf{matchHLA}(\seshce{hla_d}{\bssform}$: vector, $\seshce{ahla_r}{\bssform}$: vector) $\rightarrow$ int}\label{alg:hla_cross}
    \begin{algorithmic}[1]
        \State $\seshce{comp}{\bssform} \gets [\sesh{0}{\bssform}]^{|\ce{HLA}|}$   
        \For{$i = 0,\ldots,|\ce{HLA}|-1$} \label{cm:loop_start_vector}  \Comment{SIMD}
            \State $\seshce{comp}{\bssform}[i] \gets \sesh{\ce{hla_d}[i]}{\bssform} \land \sesh{\ce{ahla_r}[i]}{\bssform}$ \label{cm:set_entry}  
        \EndFor \label{cm:loop_end_vector}
        \State $\seshce{combined}{\bssform} \gets \cf{ORTREE}(\seshce{comp}{\bssform})[0]$
            \State\Return $\lnot \seshce{combined}{\bssform}$ \label{cm:invert_result}    
    \end{algorithmic}
\end{algorithm}

We present the main protocols for the
compatibility assessment in the following. The subprotocols for assessing the individual matching criteria HLA mismatches, ABO blood type, age, sex, and weight are given as Subprotocols~\ref{alg:hla_mux}
to~\ref{alg:size_mux} in the Appendix.

The HLA crossmatch is shown in Subprotocol~\ref{alg:hla_cross}. It tests whether
the human leukocyte antigens of the donor are unsuitable to the human leukocyte
antibodies of the recipient rendering them incompatible.

The subprotocol takes a vector with the antigens of a donor $\ce{hla}_d$ and a
vector with the antibodies of the recipient $\ce{ahla}_r$ as input. The number
of observed HLA, denoted by $|\ce{HLA}|$, is publicly known. A vector
$\ce{comp}$ stores whether the recipient possesses an antibody against any of
the donor's HLA (cf.~\lnsg{cm:set_entry}). For enhanced efficiency, we
parallelize this comparison as \emph{Single Instruction, Multiple Data} (SIMD)
operation such that all HLA matches of one patient are computed in just one
step. Afterwards, the overall compatibility (i.e., no antigen-antibody mismatch was found) is computed with an \ORGate{} gates in a tree structure, to reduce the (multiplicative) dephts of the circuit from $|\ce{HLA}|$ to $\log_2(|\ce{HLA}|)$.
To prepare for further processing, we invert $\ce{combined}$ and return it as
result of the HLA crossmatching in~\lnsg{cm:invert_result}.

\floatname{algorithm}{Protocol}
\begin{algorithm}
    \caption{$\cf{computeCompatibilityGraph}(\seshce{pairs}{\bssform}$: vector, $\seshce{w}{\assform}$: vector) $\rightarrow$ weighted adjacency matrix}\label{alg:comapt_graph_mux}
    \begin{algorithmic}[1]
        \State $\seshce{compG}{\assform} \gets matrix \in \{ \sesh{0}{\assform}\}^{|\ce{pairs}|\times |\ce{pairs}|}$
        \For{$i = 0,\ldots, |\ce{pairs}|-1$}
            \For{$j = 0, \ldots,|\ce{pairs}|-1$}
                \State $d \gets \ce{pairs}[i].d$ \Comment{Extract donor.} \label{cg:compute_edge_weight}
                \State $r \gets \ce{pairs}[j].r$ \Comment{Extract recipient.}
                \State $\seshce{edge\_w}{\assform} \gets \sesh{1}{\assform} + $\par
        \hskip\algorithmicindent $\sesh{w}{\assform}[0] \cdot \btoa(\cf{evalHLA}(\sesh{d.\ce{hla}}{\bssform}, \sesh{r.\ce{hla}}{\bssform}))+ $\par
        \hskip\algorithmicindent $\sesh{w}{\assform}[1] \cdot \btoa(\cf{evalABO}(\sesh{d.\ce{bg}}{\bssform}, \sesh{r.\ce{bg}}{\bssform})) +$\par
        \hskip\algorithmicindent $\sesh{w}{\assform}[2] \cdot \btoa(\cf{evalAge}(\sesh{d.\ce{a}}{\bssform}, \sesh{r.\ce{a}}{\bssform})) + $\par
        \hskip\algorithmicindent $\sesh{w}{\assform}[3] \cdot \btoa(\cf{evalSex}(\sesh{d.\ce{sex}}{\bssform}, \sesh{r.\ce{sex}}{\bssform})) + $\par
        \hskip\algorithmicindent $\sesh{w}{\assform}[4] \cdot \btoa(\cf{evalWeight}(\sesh{d.\ce{weight}}{\bssform}, \sesh{r.\ce{weight}}{\bssform}))$ \label{cg:compute_end}
                \State $\seshce{compG}{\assform}[i][j] \gets $\par
        \hskip\algorithmicindent $\btoa({\cf{matchHLA}(\sesh{d.\ce{hla}}{\bssform}, \sesh{r.\ce{ahla}}{\bssform}) > \sesh{0}{\bssform}}$ \textbf{ ? }\par
        \hskip\algorithmicindent ${\atob(\seshce{edge\_w}{\assform})}$\textbf{ : }${\sesh{0}{\bssform}})$   \label{cg:compatible}
            \EndFor
        \EndFor
        \State\Return $\seshce{compG}{\assform}$
    \end{algorithmic}
\end{algorithm} 

In \prot{alg:comapt_graph_mux}, we present our \mpc protocol that combines the
results of the evaluated six medical criteria influencing the compatibility of a
kidney donation into a weighted adjacency matrix indicating the donor-recipient
compatibility, named $\ce{compG}$. 

It takes a vector $\ce{pairs}$ containing all possible pairs of donors and
recipients and a vector $w$ with a weight for each criteria (i.e., how much it
influences the overall probability for good compatibility compared to the other
factors) as input. \lnpl{cg:compute_edge_weight} to~\ref{cg:compute_end}
additively combine the computed weighted probability of each compatibility
criterion and assign it to the respective edge representing the donor of the $i$-th
pair and the patient of the $j$-th pair, where $i\neq j$ and
$i,j\in\{0,\ldots,|\ce{pairs}|-1\}$. In~\lnsg{cg:compatible}, we additionally
check whether the $i$-th donor and the $j$-th patient exhibit general
immunological compatibility using the HLA crossmatch subprotocol
(cf.~Subprotocol~\ref{alg:hla_cross}). If this is the case, we store the result
of the edge weight at the respective index, otherwise we store the secret shared
constant $0$.

\noindent\emph{\mpc Cost.}
The two sections in Subprotocol~\ref{alg:hla_cross} evaluate
$|\ce{HLA}|$ \ANDGate{} gates (as SIMD) and $\log_2(|\ce{HLA}|)$ \ORGate{}\footnote{$A \lor B = 1 \oplus ((1 \oplus A) \land
(1 \oplus B))$} gates, respectively. Finally, we invert $\ce{combined}$ once.
This results in a circuit depth of $\log_2(|\ce{HLA}|)+1$ and a total number of
\ANDGate{} gates of $2\times|\ce{HLA}|$. Boolean sharing (\bss) is used in this
protocol, as Boolean operations are performed and the circuit depths is low,
thanks to the SIMD vectorization~\cite{dsz15}. 

To fully assess the matching quality (\prot{alg:comapt_graph_mux}), all
criteria have to be evaluated for each recipient,
i.e.,~\prots{alg:hla_cross},~\ref{alg:hla_mux},~\ref{alg:age_mux},~\ref{alg:abo},~\ref{alg:sex_mux},
and \ref{alg:size_mux} are run $|\ce{pairs}|^2$ times. Then, in
\prot{alg:comapt_graph_mux}, we additionally evaluate five multiplications, five
additions, one comparison, one \ANDGate{} gate, and one \MUXGate{} gate. Due to
the arithmetic operations in this protocol, the results of the compatibility
evaluation protocols must be converted between \bss~and~\ass. 

\subsubsection*{Cycle Computation}

The second phase of \ourname{} computes the number of possible kidney exchange
cycles given a concrete input cycle length\footnote{As discussed in the Related
Work, we recommend 2 to 3 to foster robustness.} from the compatible donors and recipients that were output by the compatibility matching. Our \mpc protocol for
this part is shown in~\prot{alg:det_cycles_general}.

\begin{algorithm}
    \caption{$\cf{determineNumberCycles}(\seshce{compG}{\assform}$: matrix) $\rightarrow$ number of cycles}\label{alg:det_cycles_general}
    \begin{algorithmic}
        \State $\seshce{compG}{\bssform} \gets \atob(\seshce{compG}{\assform})$ \label{nc:uG}
        \State $\seshce{uG}{\assform} \gets \cf{removeWeights}(\seshce{compG}{\bssform})$ \label{line:unweightG}
        \State $\seshce{cG}{\assform} \gets \cf{pow}(\seshce{uG}{\assform},\ce{cLen})$
        \State $\sesh{|\ce{cycles}|}{\assform} \gets \sesh{0}{\assform}$
        \For{$i=0, \ldots,|\ce{pairs}|-1$}
            \State $\sesh{|\ce{cycles}|}{\assform} \gets \sesh{|\ce{cycles}|}{\assform} + \sesh{cG}{\assform}[i][i]$
        \EndFor
       \State\Return $\sesh{|\ce{cycles}|}{\assform}$
    \end{algorithmic}
\end{algorithm}

\prot{alg:det_cycles_general} takes the secret shared weighted compatibility
graph $\ce{compG}$ as input. The desired length of cycles $\ce{cLen}$ is public.
We first compute the unweighted adjacency matrix in \lnsg{line:unweightG}
(cf.~Subprotocol~\ref{alg:unweighted_mux}, in the Appendix). For the unweighted
matrix, we compute the $ \ce{cLen}$-th power using a na\"ive
implementation\footnote{Even though exhibiting a cubic runtime complexity, this
part's performance is negligible compared to the following parts (cf.\
\fig{fig:parts-cl2}), hence, an optimization is not vital.}. The entries in this
resulting matrix indicate how many paths of length $\ce{cLen}$ start at vertex
$i$ and end at vertex $j$. For cycles, the entries are on the diagonal, as start
and end vertex are identical. Following this thought, the sum of the entries of
the diagonal is the total number of cycles with the given cycle length
$\ce{cLen}$. Note that this number contains duplicates, namely,
\enquote{congruent} cycles that are the same but were found via a different
start/end vertex.\footnote{Cycle (A, B, C) and cycle (B, C, A) are duplicates,
but cycle (C, B, A) is not.} We remove the duplicates later in
Subprotocol~\ref{alg:removeDuplicates_mux} (described in the Appendix).

\noindent \emph{\mpc Cost.}
\prot{alg:det_cycles_general} contains mostly arithmetic operations
($|\ce{pairs}|^3$ multiplications and $(|\ce{pairs}|^3-|\ce{pairs}|^2$
additions), however, the computation of the unweighted adjacency matrix is most
efficiently performed in~\bss $|\ce{pairs}|^2$ comparisons and \MUXGate{}
gates). For that reason we convert $\ce{compG}$ from~\ass to~\bss
(cf.~Line~\ref{nc:uG}) and back (in \prot{alg:unweighted_mux}).

\floatname{algorithm}{Subprotocol}
\begin{algorithm}
    \caption{$\cf{findCycles}(\seshce{compG}{\gcform}$: matrix, $\ce{cCycle}$: vector, $\seshce{allCycles}{\gcform}$: vector, $\seshce{weight}{\gcform}$: int, $\seshce{valid}{\gcform}$: int) $\rightarrow$ vector of tuples}\label{alg:find_cycles_mux}
    \begin{algorithmic}[1]
        \If{$|\ce{cCycle}| ==  \ce{cLen}$}  \label{rcr:base_begin}
            \State $\seshce{weight}{\gcform} \gets \seshce{weight}{\gcform} + \seshce{compG}{\gcform}[ \ce{cLen}-1][0]$ \label{rcr:increase_weight}
            \State $\seshce{valid}{\gcform} \gets \seshce{compG}{\gcform}[ \ce{cLen}-1][0] > \sesh{0}{\gcform}\textbf{ ? }$\par
            \hskip\algorithmicindent${\seshce{valid}{\gcform} + \sesh{1}{\gcform}}\textbf{ : }{\seshce{valid}{\gcform} + \sesh{0}{\gcform}}$ \label{rcr:final_cond_start}
            \State $\seshce{addC}{\gcform} \gets \seshce{cLen}{\gcform} == \seshce{valid}{\gcform}$ \label{rcr:final_cond_end}
            \State $\seshce{cWeight}{\gcform} \gets \tern{\seshce{addC}{\gcform}}{\seshce{weight}{\gcform}}{\sesh{0}{\gcform}}$ \label{rcr:invalid}
            \State $\seshce{cycle}{\gcform} \gets \seshce{cCycle}{\gcform}$
            \State $\seshce{allCycles}{\gcform}.\cf{append}((\seshce{cWeight}{\gcform}, \seshce{cycle}{\gcform}))$
            \State $\cf{revert}()$ \label{revert_1}
        \Else    
            \For{$i=0,\ldots, |\ce{pairs}|-1$} \label{rcr:loop_begin}
                    \If{$\ce{cCycle}.\cf{contains}(i)$}  \label{rcr:skip}
                        \State $\cf{continue}$
                    \Else
                        \State $\seshce{weight}{\gcform} \gets \seshce{weight}{\gcform} + \seshce{compG}{\gcform}[-1][i]$ \label{rcr:start_new_v}
                        \State $\seshce{valid}{\gcform} \gets \seshce{compG}{\gcform}[-1][0] > \sesh{0}{\gcform}\textbf{ ? }$\par
                        \hskip\algorithmicindent$\seshce{valid}{\gcform} + \sesh{1}{\gcform}\textbf{ : }{\seshce{valid}{\gcform} + \sesh{0}{\gcform}}$
                        \State $\ce{cCycle}.\cf{append}(i)$ \label{rcr:append}
                        \State $\seshce{allCycles}{\gcform} \gets  \cf{findCycles}(\seshce{compG}{\gcform},$\par
                        \hskip\algorithmicindent$ \ce{cCycle},\seshce{allCycles}{\gcform}, \seshce{weight}{\gcform}, \seshce{valid}{\gcform})$ 
                        \State $\ce{cCycle}.\cf{remove}()$ \label{revert_2a}
                        \State $\cf{revert}()$\label{revert_2b}
                    \EndIf
            \EndFor \label{rcr:loop_end}
        \EndIf
        \State\Return $\seshce{allCycles}{\gcform}$
    \end{algorithmic}
\end{algorithm}

\subsubsection*{Cycle Evaluation}

The third phase of \ourname{} then identifies the most likely successful unique
exchange cycles consisting of compatible pairs of donors and recipients.

Our first subprotocol for this phase, shown in
Subprotocol~\ref{alg:find_cycles_mux}, finds all exchange cycles of the desired
length (including duplicates) and computes the weight of each cycle. This weight
is the sum of all included weighted edges. As mentioned before, the weight
associated with an exchange cycle indicates the probability of the
transplantation being successfully carried out, i.e., its~robustness.

The subprotocol takes the secret shared compatibility graph $\ce{compG}$ output
by \prot{alg:comapt_graph_mux}, the currently analyzed exchange cycle
$\ce{cCycle}$, its secret shared weight $\ce{weight}$, a secret shared counter
$\ce{valid}$ which tracks the number of edges in $\ce{cCycle}$, and a vector of
secret shared tuples $\ce{allCycles}$ which will be consecutively filled with
all possible exchange cycles and the corresponding sum of weights. In a
recursive execution of Subprotocol~\ref{alg:find_cycles_mux}, this vector is
filled as explained in detail in the following. The desired output cycle length
$\ce{cLen}$ and the number of recipient-donor pairs $|\ce{pairs}|$ are public.
Additionally, also the output number of cycles $|\ce{cycles}|$ found
in~\prot{alg:det_cycles_general} is revealed for efficiency reasons. We consider
this leakage as acceptable since it leaks only a very high-level aggregate
property, generally not allowing the inference of the compatibility graph's topology\footnote{Exceptions are fully connected and unconnected graphs, as well as for $|\ce{cycles}|=1$ at pathological graph topologies. The first topologies have no security implication whatsoever and the later can, e.g., be easily avoided by introducing a check ensuring that the output is only revealed when more cycles have been found.}.

Subprotocol~\ref{alg:find_cycles_mux} first checks if the currently analyzed
exchange cycle $\ce{cCycle}$ already has the desired length $\ce{cLen}$. If this
is the case, the weight of the last edge is added to the respective sum of this
cycle's weights in \lnsg{rcr:increase_weight}. Next, each valid $\ce{cCycle}$
is added to $\ce{allCycles}$ with its respective sum of weights. A $\ce{cCycle}$
is valid, if it is closed (cf.~\lnpl{rcr:final_cond_start}
to~\ref{rcr:final_cond_end}). An invalid cycle is associated with weight zero
(cf.~\lnsg{rcr:invalid}). Note that a weight of zero does not contribute to the
solution, hence a cycle with weight zero is never considered for a solution.
In Line~\ref{revert_1}, the operations done in \lnpl{rcr:increase_weight}
to~\ref{rcr:final_cond_start} are reverted to restore the state of $\ce{cCycle}$
before the last edge was added, i.e., the weight of the last edge is subtracted
from $\ce{weight}$ and $\ce{valid}$ is decreased by $0$ (no edge) or~$1$~(edge).

Cycles that do not have the
desired length yet are handled in \lnpl{rcr:loop_begin} to~\ref{rcr:loop_end}. For these exchange cycles, the subprotocol
checks whether they are already part of $\ce{cCycle}$ as each vertex may only
appear at most once (cf.~\lnsg{rcr:skip}). If it is not included, the weight of
the edge from the previous to the new vertex is added by increasing
$\ce{cCycle}$'s weight and counter $\sesh{valid}{\gcform}$, and the new vertex is
added to $\ce{cCycle}$ (cf.~\lnpl{rcr:start_new_v} to~\ref{rcr:append}).
Afterwards, Subprotocol~\ref{alg:find_cycles_mux} is recursively called again
with the newly added vertex. Once the function returns, we revert the operations
done before to be able to analyze the next cycle~(cf.~Lines~\ref{revert_2a}
to~\ref{revert_2b}).

The second subprotocol of the cycle evaluation
(cf.~Subprotocol~\ref{alg:removeDuplicates_mux} in the \suppl) removes
duplicates from the exchange cycles set. It extracts $\#{\ce{unique}}=\lfloor
\frac{\#{\ce{cycles}}}{\ce{cLen}} \rfloor$ cycles and returns the $k$ cycles
with the highest probability for a successful~transplantation.

\floatname{algorithm}{Protocol}
\begin{algorithm}[H]
    \caption{$\cf{evaluateCycles}(\seshce{compG}{\gcform}$: matrix) $\rightarrow$ vector of tuples} \label{alg:eval_cycles}
    \begin{algorithmic}[1]
        \State $\seshce{allCycles}{\gcform},\ce{cCycle} \gets \emptyset$ 
        \label{fcc:init_start}
        \State $\seshce{weight}{\gcform},\seshce{valid}{\gcform} \gets \sesh{0}{\gcform}$
        \For{$i=0, \ldots, |\ce{pairs}|-1$}
            \State $\ce{cCycle}.\cf{append}(i)$
            \State $\seshce{allCycles}{\gcform} \gets \cf{findCycles}(\seshce{compG}{\gcform}, \ce{cCycle},$\par
            \hskip\algorithmicindent$  \seshce{allCycles}{\gcform}, \seshce{weight}{\gcform}, \seshce{valid}{\gcform})$
            \State $\ce{cCycle}.\cf{remove}()$
        \EndFor \label{fcc:init_end}
        \State $|\ce{allCycles}| \gets \cf{totalCycles}()$ 
        \State $\seshce{sortedCycles}{\gcform} \gets \cf{kNNSort}(\seshce{allCycles}{\gcform}, |\ce{cycles}|)$
        \State $|\ce{unique}| \gets \lfloor \tfrac{|\ce{cycles}|}{ \ce{cLen}} \rfloor$ \label{fcc:unique}
        \State $\seshce{filteredCycles}{\gcform} \gets \cf{removeDuplicates}(\seshce{sortedCycles}{\gcform}$) \label{fcc:rd}
        \State\Return $\seshce{filteredCycles}{\gcform}$
    \end{algorithmic}
\end{algorithm}

\prot{alg:eval_cycles} combines the previously discussed subprotocols. It first
calculates the sum of weights for each cycle with
Subprotocol~\ref{alg:find_cycles_mux} (findCycles) and sorts the result using
Subprotocol~\ref{alg:kNNSort_mux} (kNNSort) such that only the $k$ cycles with the largest
weight are output. Those are all valid cycles, possibly including duplicates.
Afterwards, the protocol removes all duplicates within the $k$ cycles.

\noindent \textit{\mpc Cost.} 
The complexity of Subprotocol~\ref{alg:find_cycles_mux} depends on the number of
pairs $|\ce{pairs}|$, $\ce{cLen}$, and the number of possible cycles
$|\ce{allCycles}|$. It is most efficient in \gc as the \MUXGate{} gates are not
independent, thus, creating a deep circuit of depth $\mathcal{O}(|\ce{allCycles}|
\times |\ce{cycles}| \times \ce{cLen})$. For removing duplicates and extracting
the most robust $k$ exchange circuits, we evaluate
$\#{\ce{cycles}}\times(\#{\ce{unique}}+\sum_{i =
0}^{\#{\ce{cycles}}}\left(\ce{cLen} \times (\ce{cLen}-1)\right)$) comparisons,
$\#{\ce{cycles}}$ $\times$ $\sum_{i = 0}^{\#{\ce{cycles}}}((\ce{cLen} \times
(\ce{cLen}-1)))$ \ANDGate{} gates, $\#{\ce{cycles}}$ $\times$ $\sum_{i =
0}^{\#{\ce{cycles}}}\left(\ce{cLen}-1\right)$ \ORGate{} gates, $\#{\ce{cycles}}$
$\times \#{\ce{unique}} \times (1 + \ce{cLen}) + \#\ce{cycles}$ \MUXGate{}
gates. This step is most efficient with \gc as the circuit is very deep. Thus, the complete cycle
evaluation routine is most efficient in \gc as each of our subroutines is most
efficient in \gc.

\subsubsection*{Solution Evaluation}
The fourth phase of \ourname{} determines the final output, a set of
\emph{disjoint} exchange cycles exhibiting the highest probability for a
successful transplantation. As a pair of donor and recipient can only be
involved in one exchange cycle, the output sets must be vertex disjoint. Note
that we find a locally optimal solution which might differ from the globally
optimal solution\footnote{Calculation of a global solution is provably a
$\mathcal{NP}$-hard problem~\cite{biroInapproximabilityKidney2007a}.}. This last
part of \ourname{} is shown in \prot{alg:find_solution_mux}.

\begin{algorithm}
    \caption{$\cf{evalSolution}(\seshce{filteredCycles}{\gcform}$: vector of tuples) $\rightarrow$ tuple(int, vector of vectors)} \label{alg:find_solution_mux}
    \begin{algorithmic}[1]
      \State $\seshce{sets}{\gcform} \gets \emptyset$ 
       \State $\seshce{weights}{\gcform} \gets \emptyset$ 
        \State $\seshce{dummyC}{\gcform} \gets \{\sesh{|\ce{pairs}|}{\gcform}\}^{ \ce{cLen}}$ \label{fs:result_weight}
        \For{$i = 0, \ldots, |\ce{unique}| - 1$}
            \State $\seshce{tempSet}{\gcform} \gets \emptyset $ \label{fs:init_current}
            \State $\seshce{tempSet}{\gcform}.\cf{append}(\seshce{filteredCycles}{\gcform}[i][1])$  
            \State $\seshce{weight}{\gcform} \gets \seshce{filteredCycles}{\gcform}[i][0]$ 
            \State $counter \gets 1$  \label{fs:init_end}
            \For{$j = 0, \ldots, |\ce{unique}| -1$} 
                \If{$i == j$}
                    \State $continue$
                \EndIf
                \State $\seshce{cCycle}{\gcform} \gets \seshce{filteredCycles}{\gcform}[j][1]$
                \State $\seshce{disjoint}{\gcform} \gets \cf{disjointSet}(\seshce{tempSet}{\gcform},$\par
                \hskip\algorithmicindent$ \seshce{cCycle}{\gcform})$ \label{esg:find_disjoint_set}
                \State $\seshce{vertices}{\gcform} \gets \emptyset$
                \State $\seshce{vertices}{\gcform}.\cf{append}({\seshce{disjoint}{\gcform}}$\textbf{ ? }\par
            \hskip\algorithmicindent$\seshce{cCycle}{\gcform}\textbf{ : }{\seshce{dummyC}{\gcform}})$  \label{fs:add_Begin}
                \State $\seshce{weight}{\gcform} \gets \tern{\seshce{disjoint}{\gcform}}{\seshce{weight}{\gcform}}{\sesh{0}{\gcform}}$
                \State $\seshce{tempSet}{\gcform}.\cf{append}(\seshce{vertices}{\gcform})$
                \State $\ce{counter} \gets \ce{counter} + 1$ \label{fs:add_end}
            \EndFor
            \State $\seshce{sets}{\gcform}.\cf{append}(\seshce{tempSet}{\gcform})$  \label{fs:add_set}
            \State $\seshce{weights}{\gcform}.\cf{append}(\seshce{weight}{\gcform})$  \label{fs:add_set_weight}
       \EndFor
       \State\Return $\cf{findMaximumSet}(\seshce{sets}{\gcform}, \seshce{weights}{\gcform})$ \label{fs:result}
    \end{algorithmic}
\end{algorithm}

\prot{alg:find_solution_mux} takes a secret shared vector of tuples
$\ce{filteredCycles}$ with all valid unique cycles and their respective weights,
the number of valid cycles $|\ce{unique}|$, and the cycle length $\ce{cLen}$ as
input. The number of pairs $|\ce{pairs}|$ is a public variable as before. 

It checks each valid cycle $\ce{cCycle}$ if it is disjoint from all other
previously analyzed cycles in $\ce{tempSet}$. The \mpc subprotocol for testing
the disjointness is given in Subprotocol~\ref{alg:disjoint} in the \suppl. If
it is disjoint, $\ce{cCycle}$ is added to the set of potential solutions (Lines
\ref{fs:add_Begin} - \ref{fs:add_set_weight}). Finally, the set with the highest
weight is returned. Details of the corresponding \mpc protocol can be found in
Subprotocol~\ref{alg:max_set_mux} in the \suppl.

\noindent\emph{\mpc Cost.}
In total, we evaluate $|\ce{unique}|^2$ \ADDGate{} gates, $|\ce{unique}|^2
\times \ce{cLen}^2 + |\ce{unique}|$ comparisons, $4 \times |\ce{unique}|^2 +
|\ce{unique}|$ \MUXGate{}, and $|\ce{unique}|^2 \times \ce{cLen^2}$ \ORGate{}
gates. The solution evaluation is most efficient in \gc as there are only few
arithmetic operations and mostly comparisons.

\subsection*{Complexity Assessment}
In Table \ref{tab:complexity}, the asymptotic complexities for the four phases of SPIKE are given.

 \begin{table*}
        \caption{Complexity Assessment}
        \begin{tabular}{l|l|l|c}
            \toprule
            Phase&Name&Protocol&Time Complexity\\
            \midrule
                Part 1)&Compatibility Matching&Subprotocol~\ref{alg:hla_cross} & $\mathcal{O}(|\ce{HLA}|)$ \\
                &&Subprotocol~\ref{alg:hla_mux} & $\mathcal{O}(|\ce{HLA}|)$ \\
                &&Subprotocol~\ref{alg:abo} & $\mathcal{O}(1)$ \\
                &&Subprotocol~\ref{alg:age_mux} & $\mathcal{O}(1)$ \\
                &&Subprotocol~\ref{alg:sex_mux} & $\mathcal{O}(1)$ \\
                &&Subprotocol~\ref{alg:size_mux} & $\mathcal{O}(1)$ \\
                &&Protocol~\ref{alg:comapt_graph_mux} & $\mathcal{O}(|\ce{pairs}|^2 \times |\ce{HLA}|)$ \\
                \hline
                Part 2) &Cycle Computation&Subprotocol~\ref{alg:unweighted_mux} & $\mathcal{O}(|\ce{pairs}|^2)$ \\
                && Protocol~\ref{alg:det_cycles_general} & $\mathcal{O}(\ce{cLen} \times |\ce{pairs}|^{3})$ \\
                \hline 
                Part 3)&Cycle Evaluation&Subprotocol~\ref{alg:total_cycles} & $\mathcal{O}(1)$ \\
                &&Subprotocol~\ref{alg:find_cycles_mux} & $\mathcal{O}(|\ce{pairs}|^{\ce{cLen}})$ \\
                &&Subprotocol\ref{alg:kNNSort_mux} & $\mathcal{O}(|\ce{cyclesSet}| \times \ce{k} \times \ce{cLen})$ \\
                &&Subprotocol~\ref{alg:removeDuplicates_mux} & $\mathcal{O}(|\ce{cycles}|^2)$ \\
                &&Protocol~\ref{alg:eval_cycles} & $\mathcal{O}(|\ce{pairs}|^{\ce{cLen}})$ \\
                \hline
                Part 4)&Solution Evaluation&Subprotocol~\ref{alg:disjoint} & $\mathcal{O}(|\ce{cycles}|\times \ce{cLen})$ \\
                &&Subprotocol~\ref{alg:max_set_mux} & $\mathcal{O}(\ce{cycles}|^{2})$ \\
                && Protocol~\ref{alg:find_solution_mux} & $\mathcal{O}(|\ce{cycles}|^{3} \times \ce{cLen^2})$ \\
            \bottomrule
        \end{tabular}
        
        \label{tab:complexity}
\end{table*}

The most important parameters of the first part, the Compatibility Matching shown in the first section of the table, are the
number of HLA (cf. Background) $|{\ce{HLA}}|$ and the number of pairs $|{\ce{pairs}}|$. In the default configuration, $|\ce{HLA}|$ is $50$.
For the second phase, the dominant parameter is the number of
pairs $|\ce{pairs}|$.
In the third section of \tab{tab:complexity}, the asymptotic complexity for the Cycle Evaluation is given. The relevant parameters here are
the number of pairs $|\ce{pairs}|$, the total number of cycles $|\ce{allCycles}|
= |\ce{pairs}|^{\ce{cLen}}$, the number of existing cycles $|\ce{cycles}|$, the
number of unique cycles $|\ce{unique}| = \lfloor
\tfrac{|\ce{cycles}|}{\ce{cLen}} \rfloor$, the length of cycles $\ce{cLen}$, and
the factor $k$ (i.e, the number of cycles with highest probability for successful transplantation), and the number of elements in $\ce{cyclesSet}, |\ce{cyclesSet}|$
of Protocol~\ref{alg:kNNSort_mux}.
The most important
parameters of the last phase, the Solution Evaluation, are the number of unique cycles $|\ce{cycles}|$, and the length of
cycles $\ce{cLen}$.

Overall, the asymptotic complexity of \ourname{} is:
\[
    \mathcal{O}(|\ce{pairs}|^2 \times |\ce{HLA}| + \ce{cLen} \times |\ce{pairs}|^{3} + |\ce{cycles}|^3 \times \ce{cLen}^2).
\]
The most most important parameters are the number of pairs $|\ce{pairs}|$, the
number of considered HLA $|\ce{HLA}|$, the length of cycles $\ce{cLen}$, and the
number of unique cycles $|\ce{cycles}|$.

\section*{Results}
\label{sec:results}
All benchmarks were run on two servers equipped with Intel Core i9-7960X
processors and \SI{128}{\giga\byte} RAM. They are connected via
\SI{10}{\giga \bit/\second} LAN with a median latency of \SI{1.3}{\milli
\second}. All benchmarks are averaged over 10~runs.

\subsection*{Network Setups}\label{sub:network}
To provide meaningful performance benchmarks for a variety of real-world
settings, we envision two network settings for privacy-preserving KEP that we
describe in the following. In addition, for the comparison to the works of
Breuer at al.~\cite{breuer20, breuer2022}, we replicated their network setting
with \SI{1}{\giga b/\second} bandwidth and \SI{1}{\milli \second} of latency.

\noindent \textit{LAN.}
The high-bandwidth, low latency network scenario, here referred to as
\emph{LAN}, is the most relevant real-world scenario for our application. In
Germany, most (larger) medical institutions utilize high-bandwidth Internet
connections. In the case of most university hospitals the German Research
Network (\enquote{Deutsches Forschungsnetz} DFN\footnote{\url{https://dfn.de/}}) provides dedicated, high
bandwidth communication networks. Our \emph{LAN} benchmarks are performed using
a \SI{10}{\giga b/\second} connection with an average latency of \SI{1.3}{\milli
\second}.

\noindent \textit{WAN.}
One benefit of a \mpc-based privacy-preserving KEP solution could be reduced legal and
regulatory data protection requirements, due to the high security level of the
computation itself. This would allow smaller, local hospitals and medical
practices to directly participate in the kidney exchange. Those institutions
might be connected via residential Internet access. For that scenario, we
benchmarked \ourname{} in a reduced-bandwidth, high latency network. A bandwidth
restriction to \SI{100}{\mega b/\second} with added latency of \SI{100}{\milli
\second} was implemented using the \texttt{tc}\footnotemark command to simulate
the \emph{WAN} network. The high latency was chosen to take packet loss due to
unreliable connections into account.
\footnotetext{\url{https://man7.org/linux/man-pages/man8/tc.8.html}}

\subsection*{Performance Benchmarks}\label{sub:performance}
\begin{figure*}[htbp]
  \includegraphics{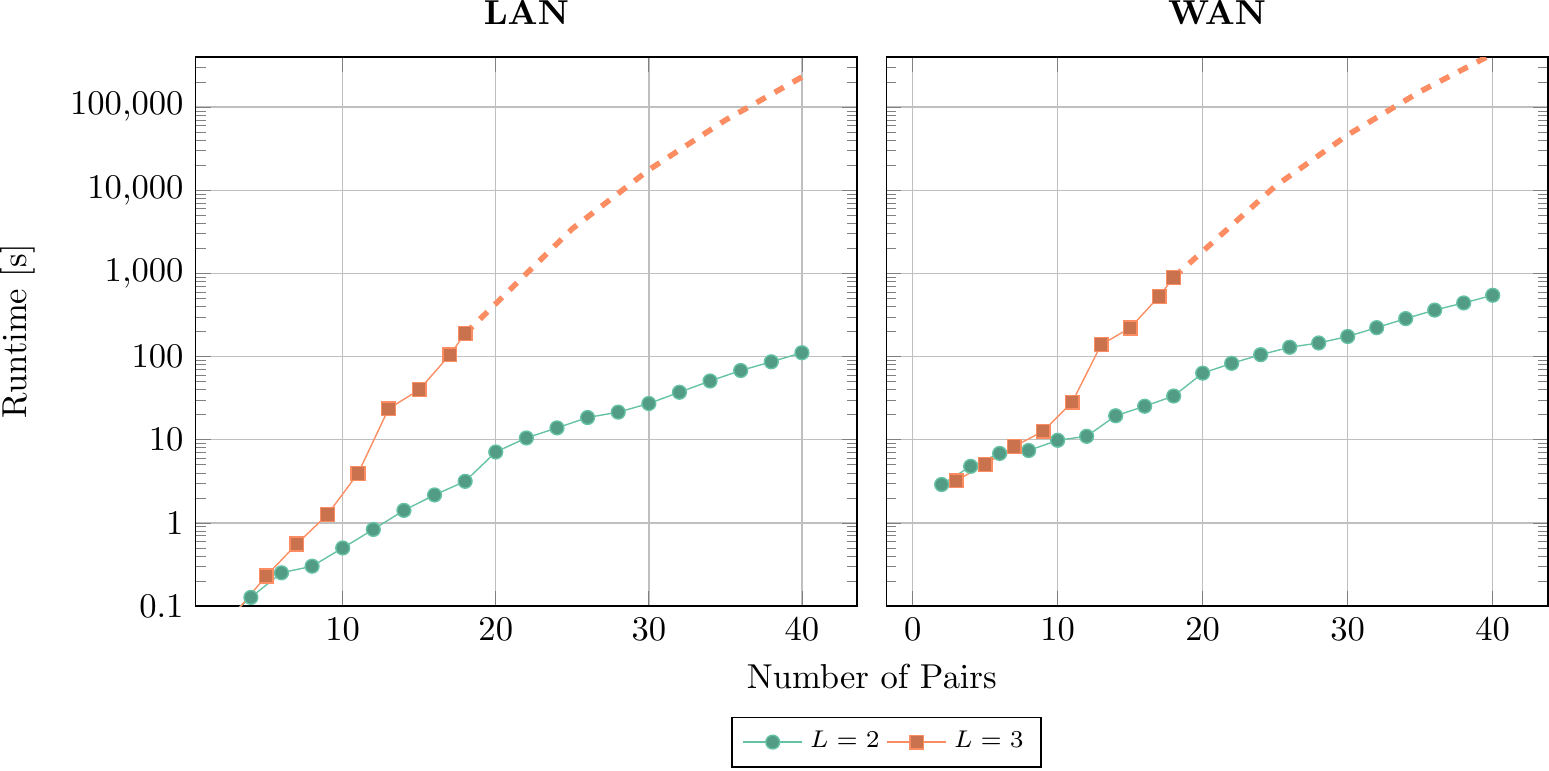}
  \caption{Overall runtime of \ourname{} for cycle lengths $L=2$ and $L=3$ in
  both network scenarios. The dashed line shows the extrapolated power function
for $L=3$.}%
  \label{fig:overall}
\end{figure*}

\fig{fig:overall} shows the total runtime of \ourname{} for varying
numbers of pairs, both network settings, and cycle lengths $L=2$ and $L=3$.
The full results are in \tabs{tab:comparison-cl2-p1}~to~\ref{tab:comparison-matching}
in the Appendix.

During the evaluation of longer cycles ($L\geq3$) RAM utilization proved itself
to be a bottleneck for execution. For those scenarios, we benchmarked up to RAM
exhaustion and extrapolated the runtimes according to the underlying power-law
complexity. The extrapolation is shown with a dashed line. The sudden increase
in runtime for $L=3$ between \num{12} and \num{13} pairs occurs due to
swapping.

As a general result, the expected polynomial relationship between the number of
pairs and the overall runtime can be observed, reflected in the power-law
development in the semilog graphs. For $L=2$, we achieve a total runtime of under
\SI{4}{\minute} for \num{40} pairs, thus, demonstrating real-world applicable
performance. The WAN setting increases the overall runtime by less than an order
of magnitude. Calculation times under \SI{20}{\minute} for \num{40} pairs in
this setting render the participation feasible for physicians with residential
Internet connections. To find a solution for larger cycle lengths, the exponent
in the time complexity increases, increasing the runtimes significantly. But even
then \num{25} pairs are computable in around \SI{1}{\hour}.

\begin{figure*}[htbp]
  \includegraphics{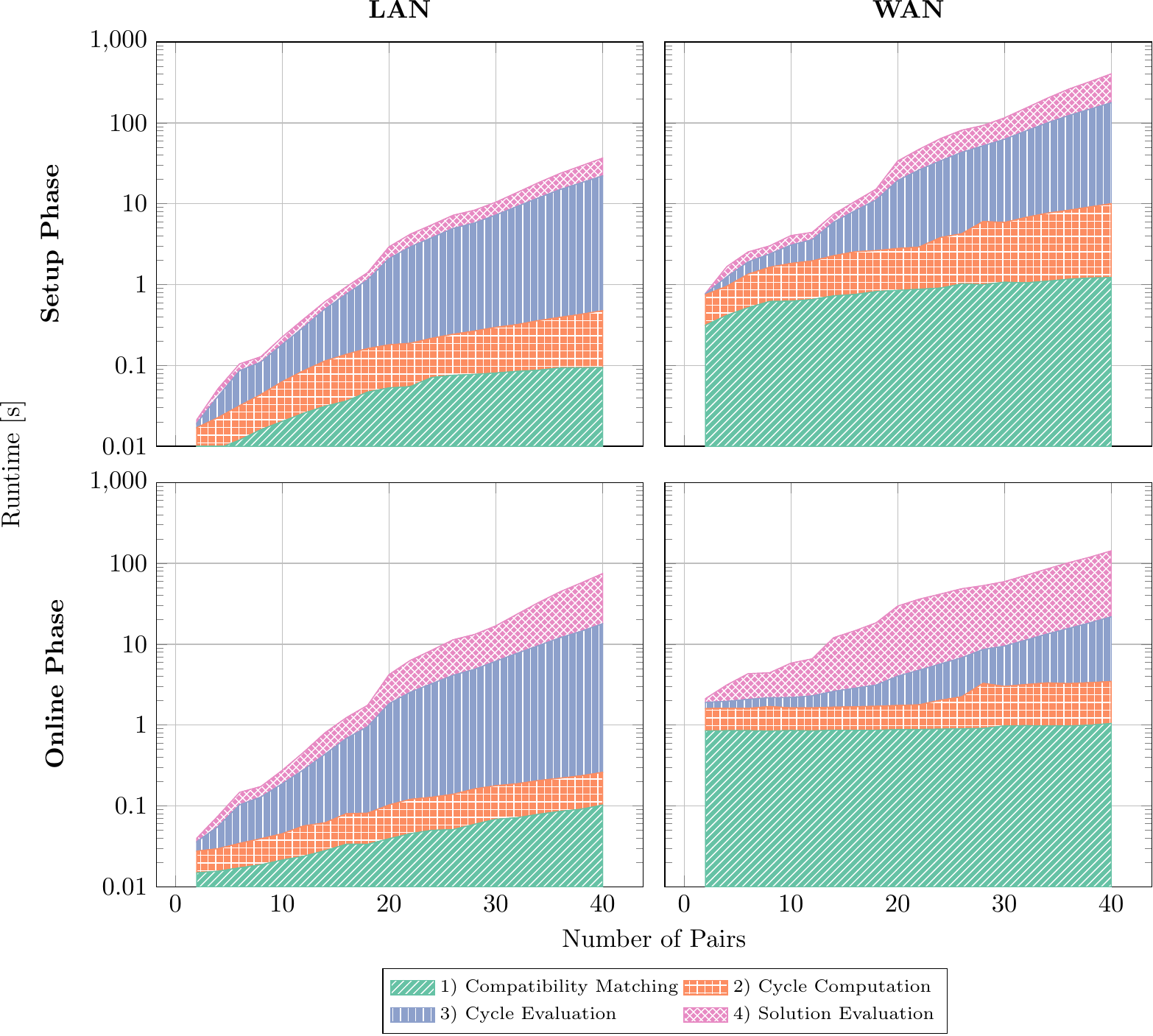}
  \caption{Runtime of \ourname{} for $L=2$ separated by algorithmic
  parts, protocol phase, and network setting.}%
  \label{fig:parts-cl2}
\end{figure*}

\fig{fig:parts-cl2} shows the runtimes of the individual parts of the algorithm
($L=2$). It is clearly visible, that the medical compatibility testing and graph
creation, as well as the cycle computation quickly become negligible compared to
the runtimes of cycle evaluation and the evaluation of the global solution. The
duration of online and offline phases are in the same order of magnitude. By
executing the phases separately, a \SI{134}{\percent} performance increase in
the online execution can be achieved, compared to the accumulated runtime
(cf. \fig{fig:overall}).

\subsection*{Comparison to State-Of-The-Art}\label{sub:comparison}

\begin{figure}[htbp]
  \includegraphics{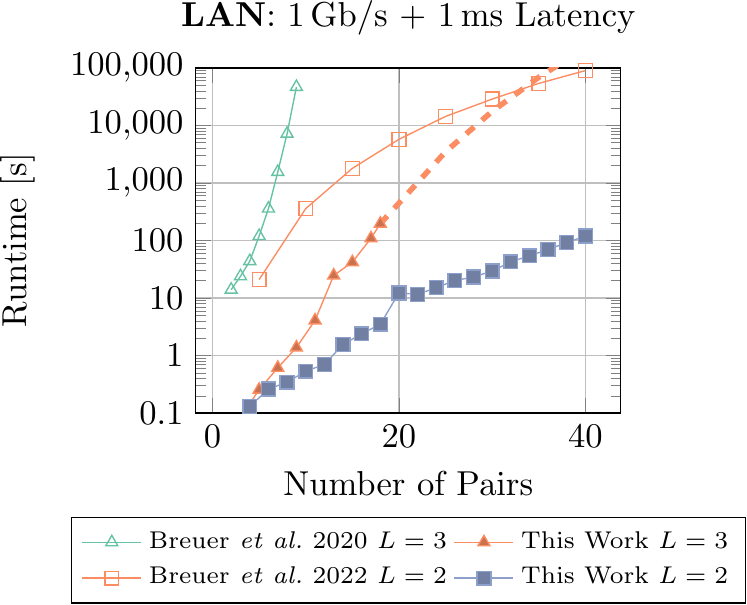}
  \caption{Runtime comparison between this work with cycle lengths $L=2$ and
    $L=3$, and both Breuer \textit{et al.} 2020 ($L=3$)~\cite{breuer20} and Breuer \textit{et
    al.} 2022 ($L=2$)~\cite{breuer2022}. All measurements use a LAN network setting with
  \SI{1}{\giga b/\second} bandwidth and \SI{1}{\milli \second} latency. The
dashed line shows the extrapolated power function for our algorithm at $L=3$}%
  \label{fig:comparison}
\end{figure}

In \fig{fig:comparison}, we compare the runtime of our implementation for $L=2$
and $L=3$ with two implementations from Breuer \textit{et al.}~\cite{breuer20,
breuer2022}. The first implementation~\cite{breuer20} uses a Threshold
Homomorphic Encryption scheme and enables to solve the privacy-preserving KEP
with arbitrary cycle length as in \ourname{}. The maximum cycle size is set to
$L=3$ in their benchmarks. The second one~\cite{breuer2022} is based on
three-party honest majority Shamir's Secret Sharing using the \spdz framework and limits its
cycle length to $L=2$. The performance data for both implementations is taken
from the referenced publications.

\begin{figure*}[htbp]
  \includegraphics{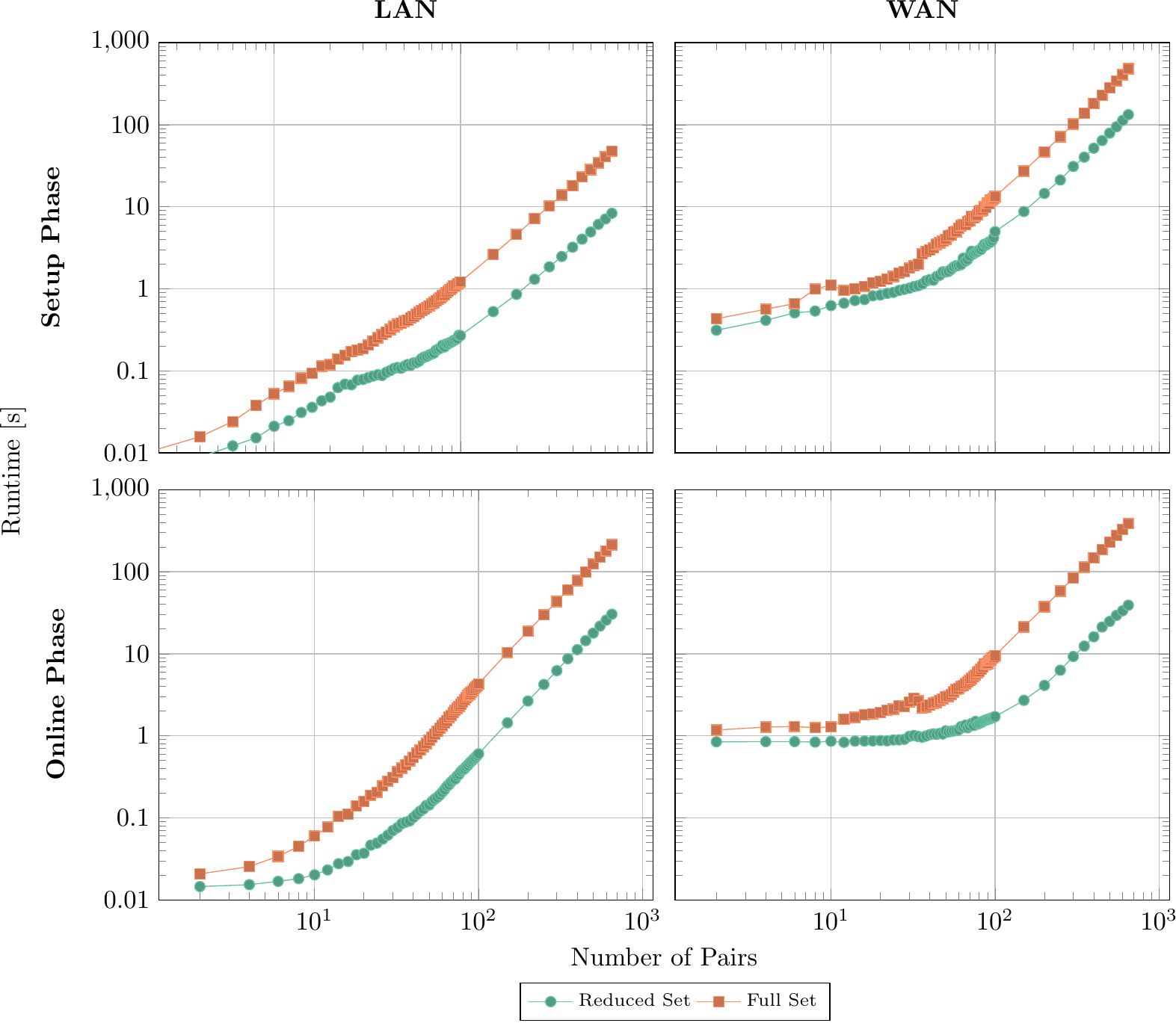}
  \caption{Comparison of the compatibility matching (i.e., compatibility graph
    generation) performance between the reduced set of criteria (Breuer
  \textit{et al.} 2020)~\cite{breuer20} and the full set of this work for cycle length
$L=2$. The runtimes of the remaining algorithmic parts are independent of this
choice.}%
  \label{fig:comparison_match}
\end{figure*}

Our implementation, as well as the \spdz based state-of-the-art~\cite{breuer2022}, shows a
polynomial-bound power-law graph. The \he-based implementation shows clearly an
exponential runtime development, increasing rapidly. For \num{9} pairs, the
maximum number of pairs benchmarked in the original publication~\cite{breuer20}, our
implementation achieves a $\num{29828} \times$ speedup. For $L=2$, our implementation
performs $\num{414} \times$ better than the \spdz-based implementation~\cite{breuer2022}.
To improve the medical quality of the donor-recipient matching, we implemented
additional matching criteria, as described in the \enquote{Background}
section. As we have seen in \fig{fig:parts-cl2}, the performance impact of the
compatibility matching algorithm is negligible compared to the runtime of the
remaining algorithmic parts. However, in~\fig{fig:comparison_match} we compare
the performance difference between the reduced set of medical matching criteria
and the full set. For small number of pairs there is a transient phase, where
the runtime of the full set rises faster. After this transient phase, both
curves assume nearly the same slope. In the plots for the WAN network model, the
latency-induced \enquote{baseline} runtime can be observed.

\section*{Discussion}%
\label{sec:discussion}

\subsection*{Security Guarantees}
Our privacy-preserving kidney exchange protocol, \ourname, is implemented using
the ABY~\cite{dsz15} \mpc framework, guaranteeing computational semi-honest
security in a two-party setting. An adversary~$\mathcal{A}$ can
corrupt at most one of the two computing parties. $\mathcal{A}$ is
assumed to follow the protocol specification and gets access to all
messages of the corrupted party (sent and received) while trying to extract private
information. This security model is standard in the privacy research community
and protects against two security concerns: 1.\ inadvertent disclosure of
sensitive data and 2.\ \emph{full} data disclosure in case of a breach in one of
the parties (in comparison to a centralized computation). In the outsourcing
scenario with two computation parties and an arbitrary number of data sources,
both computation parties \emph{must not} collude. However, an arbitrary number
of data sources is allowed to collude or behave maliciously, without breaking
the security guarantees. Note, that \mpc only gives privacy guarantees for the
computation, whereas maliciously formed inputs might lead to incorrect outputs. For a
\enquote{holistic} data privacy perspective, please see~\cite{desai2016five}.

While this adversarial model is not sufficient for all applications, e.g.,
computations with parties in different
jurisdictions~\cite{europeandataprotectionboardRecommendations012021}, it suits
our intended setting, namely the joint computation among large, intra-national
medical institutions. Both semi-honest behaviour as well as the non-collusion
assumption can be enforced by legal and regulatory means.

For a full description of the cryptographic assumptions and guarantees inherited
by the primitives used in ABY, we refer to the respective section in the
Appendix and the original ABY publication~\cite{dsz15}.

While all s' data, including the association to the various data sources
are considered to be private data and are protected by the aforementioned
guarantees, we consider the \emph{number} of donor-recipient pairs, as well as
the maximum number of cycles in the graph as public information. This choice has
important performance impacts, however, if the numbers of pairs are to be
considered private as well, the real numbers can be hidden by padding each input
array to a fixed length with dummy entries.

\subsection*{Real-World Deployability}
This work introduces a protocol for finding a solution for the kidney exchange
problem in a privacy-preserving fashion. As demonstrated in the performance
benchmarks and the security discussion, it meets all initially determined
requirements for a secure privacy-preserving KEP w.r.t privacy, efficiency,
decentralization, and adaptability for medical experts.

Concretely, it enables real-world periodic batch-processing of a large number of
donor-recipient pairs and a practical cycle length of $L=3$, even in residential
network settings. This allows even residential nephrology experts to participate
in kidney exchanges, hence, providing a better medical care for their patients.
However, \ourname{} requires a significant amount of communication, thus, it is
not yet ready for the usage of metered or cell data connections in a two-party computation protocol\footnote{In contrast, \ourname{} is already practical for an outsourcing scenario where mobile clients secret share their data among two non-colluding servers or cloud entities.} which might be
an interesting direction for future work.

By using state-of-the-art provably secure cryptographic techniques, the privacy
of sensitive medical information of donors and recipients is fully protected by
clearly defined hardness assumptions of mathematical problems. Furthermore, by
pursuing a completely decentralized approach without a trusted third party, the
risk of data leakages in case of a data security incident at one participating
facility is significantly reduced, compared to a breach in a central computation
node or repository.

Allowing medical professionals to choose many parameters of the algorithm to adapt
to new evidence-based guidelines or specific situational constraints ensures
flexibility and maintainability for future application. The compatibility
matching algorithm is configurable by choosing the considered HLA, as well as
the weights of the chosen medical factors. This explicitly allows the
deactivation of chosen comparisons. Due to the clear architecture boundaries in
the open source implementation, additional checks and criteria can easily be
included.

While meeting all formal requirements, \ourname{} falls short in two aspects:
first, we observe a high memory consumption during the computation. This is
expected, as this protocol was optimized for runtime performance. Additionally,
hardware costs are, generally speaking, no hindrance for meeting data protection
regulation. Thus, this aspect does not jeopardize the adoption in the intended
use cases. However, developing internal batch processing of graph clusters and
the employment of space-optimized data structures might be worthwhile
opportunities for further research. An interesting direction for future work can
be to explore the compatibility with recent advances in \mpc-based graph
analysis for breadth-first search~\cite{araki2021} scaling linearly in the
number of vertices. Second, the developed software components are research
artifacts and fulfil a prototypical function. For real-world adoption the
implementation of widespread medical standards, e.g., HL7 FHIRE
R4\footnote{\url{https://www.hl7.org/fhir/R4/}}, audit- and authentication
capabilities, integration in medical research pipelines, creation of deployment
packages, and lastly full (legal) documentation must be pursued. This is,
however, not in the scope of this work.

\section*{Conclusion}
In this work, we introduced \ourname, an efficient privacy-preserving
Kidney Exchange Problem (KEP) protocol. Using provably secure cryptographic
techniques, \ourname{} provides highest data protection guarantees for
patients' sensitive medical data, while allowing a decentralized computation of
a solution to the kidney exchange problem. We implement adaptable medical
compatibility matching algorithms, giving medical professionals the flexibility
to accommodate updated guidelines and the specific situational constraints.

Our optimized protocols achieve a $\num{30000}\times$ and $\num{400}\times$
speedup compared to the current state-of-the-art~\cite{breuer20,breuer2022} for
cycle lengths of $L=3$ and $L=2$, respectively. With a total runtime of under
\SI{4}{\minute} for \num{40} pairs at $L=2$ and around \SI{1}{\hour} for
\num{25} pairs at $L=3$, we demonstrate feasible performance for many real-world
applications.

We hope that the advancements in privacy protection and application
performance will allow more medical facilities to participate in Kidney
Exchanges, thus increasing the recipients' chances for timely and potentially
live-saving surgery.


\begin{backmatter}

\section*{Acknowledgements}
Many thanks to Ulrich
Zwirner for sharing his medical knowledge during our fruitful discussions.

\section*{Funding}
This project received funding from the European Research Council~(ERC) under the
European Union's Horizon 2020 research and innovation program~(grant agreement
No.850990 PSOTI). It was co-funded by the Deutsche Forschungsgemeinschaft(DFG)
-- SFB1119 CROSSING/236615297 and GRK2050 Privacy \& Trust/251805230, by the
German Federal Ministry of Education and Research and the Hessen State Ministry
for Higher Education, Research and the Arts within ATHENE, and by the German
Federal Ministry of Education and Research within the HiGHmed project
(\#01ZZ1802G).

\section*{Abbreviations}
\mpc: Secure Multi-Party Computation; HE: Homomorphic Encryption; SSS: Shamir's Secret Sharing; \gc: Yao's Garbled Circuits; \bss: Boolean Secret Sharing; \ass: Arithmetic Secret Sharing; PPKE: Privacy-Preserving Kidney Exchange; KEP: Kidney Exchange Problem; HLA: Human
Leukocyte Antigens

\section*{Availability of data and materials}
Our code and the datasets analyzed are available here:~\url{https://encrypto.de/code/PPKE}.

\section*{Ethics approval and consent to participate}
Not applicable.

\section*{Competing interests}
The authors declare that they have no competing interests.

\section*{Consent for publication}
Not applicable.

\section*{Authors' contributions}
TB implemented the discussed software.\\
TB and TK specified the medical requirements.\\
HM and TK designed the research project and guided the protocol design.\\
TB, TK, and HM designed the performance benchmarks which were conducted and analyzed by TB and TK.\\
TS and KH led this research project.\\
All authors contributed to the manuscript and substantively revised it. All
authors read and approved the final manuscript.



\bibliographystyle{bmc-mathphys} 
\bibliography{bmc_article}      







\section*{Appendix}


\subsection*{Additional \mpc Subprotocols}\label{app:add_protocols}
This part of the Appendix contains additional protocols of our \mpc-based PPKE solution, \ourname, that are similar to the ones presented in the main part and, thus, referred to the appendix.
\subsubsection*{Additional Protocols for the Compatibility Matching}
This subsection presents additional subprotocols for the compatibility matching phase.
\floatname{algorithm}{Subprotocol}
\begin{algorithm}
    \caption{$\cf{evalHLA}(\sesh{\ce{hla_d}}{\bssform}$: vector,$\sesh{\ce{hla_r}}{\bssform}$: vector) $\rightarrow$ int}\label{alg:hla_mux}
    \begin{algorithmic}[1]
        \State $\sesh{\ce{mm}}{\bssform} \gets \{\sesh{0}{\bssform}\}^{|\ce{HLA}|}$
        \For{$i = 0,\ldots, |\ce{HLA}| - 1$}  \label{hla:loop_start}
            \State $\seshce{mm}{\bssform} \gets \seshce{hla_d}{\bssform}[i] \oplus \seshce{hla_r}{\bssform}$ \label{hla:mm}\Comment{SIMD} 
        \EndFor \label{hla:loop_end}
        \State $\seshce{sum}{\bssform} \gets \cf{HammingW}(\{\sesh{0}{\bssform}\}^{|\ce{HLA}|}, \seshce{mm}{\bssform})$ \label{hla:line_hamming_distance}
        \State $\seshce{c}{\bssform} \gets \seshce{sum}{\bssform} < \sesh{5}{\bssform}$ \label{hla:begin_select}
        \State $\seshce{b}{\bssform} \gets \seshce{sum}{\bssform} < \sesh{3}{\bssform}$
        \State $\seshce{a}{\bssform} \gets \seshce{sum}{\bssform} == \sesh{0}{\bssform}$ 
        \State\Return ${\seshce{a}{\bssform}}$\textbf{ ? }\par
        \hskip\algorithmicindent${\seshce{A}{\bssform}}{\left( \tern{\seshce{b}{\bssform} }{\seshce{B}{\bssform}}{\left(\tern{\seshce{c}{\bssform}}{\seshce{C}{\bssform}}{\sesh{0}{\bssform}}\right)} \right)}$ \label{hla:end_select}
    \end{algorithmic}
\end{algorithm}

\noindent\textbf{HLA Antigen Comparison.} In \sprot{alg:hla_mux}, we compare the HLA antigens of the potential recipient and donor and determine the number of HLA mismatches. It takes two vectors $\ce{hla_d}$ and $\ce{hla_{r}}$ with the HLA antigens of the donor and recipient respectively as input. The number of $|\ce{HLA}|$ is public as it is a fixed value. The vector $\ce{mm}$ indicates the HLA mismatches of the donor and the recipient. A mismatch occurs if either donor or recipient has a HLA antigen that the other does not have (cf.~\lnsg{hla:mm}). For enhanced efficiency, we parallelize the comparison as SIMD operation such that the vector $\ce{mm}$ is computed in a single step. Afterwards, the number of HLA mismatches is determined with a Hamming Weight Circuit~(cf.~Line~\ref{hla:line_hamming_distance}). Based on the number of mismatches, the subprotocol outputs an indicator for the quality of the pairing w.r.t. the HLA antigens: Class $A$ is an optimal fit with no mismatches, class $B$ is a good fit, and class $C$ is an acceptable fit with 3-4 mismatches (cf.~Lines~\ref{hla:begin_select}-\ref{hla:end_select}).

\noindent\emph{\mpc Cost.}
\lnsg{hla:mm} in \sprot{alg:hla_mux} evaluates $|\ce{HLA}|\times$\XORGate{} gates (as SIMD). \lnsg{hla:line_hamming_distance} evaluates one Hamming Distance circuit. \lnpl{hla:begin_select}-\ref{hla:end_select} contain three comparison and three \MUXGate{} gates. Thus, the circuit's multiplicative depth is $7$, which is determined by the number of \ANDGate{} gates on the longest path. Naively, using Yao's Garbled Circuits (\gc) seems to be most efficient. However, considering that \sprot{alg:hla_cross} is done in \bss sharing, the conversion cost outweigh the benefits of using \gc instead of \bss, thus, \bss is used here as well.

\begin{algorithm}
    \caption{$\cf{evalABO}(\seshce{bg_d}{\bssform}$ : vector, $\seshce{bg_r}{\bssform}$ : vector) $\rightarrow$ int}\label{alg:abo}
    \begin{algorithmic}[1]
        \State $\seshce{a}{\bssform} \gets \lnot \Bigl( \bigl(\seshce{bg_r}{\bssform}[0] \oplus \seshce{bg_d}{\bssform}[0] \bigr) \lor \bigl(\seshce{bg_r}{\bssform}[1] \oplus \seshce{bg_d}{\bssform}[1] \bigr) \Bigr)$ \label{abo:case1}
        \State $\seshce{b}{\bssform} \gets \bigl(\seshce{bg_r}{\bssform}[1] \land \lnot \seshce{bg_d}{\bssform}[0] \bigr) \lor \bigl(\seshce{bg_r}{\bssform}[0] \land \lnot \seshce{bg_d}{\bssform}[1]\bigr)$ \label{abo:case2}
        \State $\seshce{v}{\bssform} \gets 
        \seshce{a}{\bssform} \lor \seshce{b}{\bssform} $
        \State\Return \tern{$\seshce{v}{\bssform}$}{$\seshce{best_{age}}{\bssform}$}{$\sesh{0}{\bssform}$}
    \end{algorithmic}
\end{algorithm}

\noindent\textbf{ABO blood group comparison. }Subprotocol~\ref{alg:abo} contains the privacy-preserving evaluation of the compatibility of ABO blood groups of a donor and a recipient. It takes two two-bit vectors as input: $bg_d\in\{0,1\}^2$ is the blood group of the donor and $bg_r\in\{0,1\}^2$ is the blood group of the recipient. The blood group encoding is shown in~\tab{tab:ABO-Encoding}. \lnpl{abo:case1}-\ref{abo:case2} ensure that the blood group of recipient and donor are compatible, i.e., they have to be either equal, $bg_r[1]>bg_d[0]$, or $bg_r[0]>bg_d[1]$ (cf.~\tab{tab:abo_comp}).

\begin{table}
        \centering
        \caption{Encoding of the different blood groups.}
        \begin{tabular}{c|c}
            \toprule
            Encoding & Blood Group  \\
            \midrule
            00 & O \\
            01 & A \\
            10 & B \\
            11 & AB \\
            \bottomrule
        \end{tabular}
        \label{tab:ABO-Encoding}
\end{table}

\noindent\emph{\mpc Cost.} Here, we evaluate 14 XOR gates and five $AND$ gates in total per donor/recipient pair. As \XORGate{} gates can be locally evaluated, they are ``for free''. Therefore, the \ANDGate{} gates and circuit depth determine which \mpc protocol is most efficient. \bss is slightly more efficient than \gc since the circuit depth is smaller than the number of total AND gates.

\begin{algorithm}
    \caption{$\cf{evalAge}(\seshce{a_d}{\bssform}$: int, $\seshce{a_r}{\bssform}$: int) $\rightarrow$ int}\label{alg:age_mux}
    \begin{algorithmic}[1]
        \State $\seshce{eq}{\bssform} \gets \seshce{a_d}{\bssform} == \seshce{a_r}{\bssform}$ \label{age:same} 
        \State $\seshce{yg}{\bssform} \gets \lnot \seshce{a_d}{\bssform} \land \seshce{a_r}{\bssform}$ \label{age:ydor}
        \State\Return $\seshce{yg}{\bssform} $\textbf{ ? }\par
        \hskip\algorithmicindent${\left(\tern{\seshce{eq}{\bssform}}{\seshce{A}{\bssform}}{\seshce{B}{\bssform}} \right)}$\textbf{ : }${\left(\tern{\seshce{eq}{\bssform}}{\seshce{A}{\bssform}}{\sesh{0}{\bssform}}\right)}$ \label{age:mux}
    \end{algorithmic}
\end{algorithm}

\noindent\textbf{Age Comparison.} \sprot{alg:age_mux} evaluates the compatibility of a donor and recipient based on their age group. It takes the age group of the donor $\langle a_d\rangle^B$ and the age group of the recipient $\langle a_r\rangle^B$ as input. \lnsg{age:same} checks if they are in the same age group and \lnsg{age:ydor} evaluates whether the donor is in a younger age group than the recipient. Afterwards, we compute the respective weight of this donor and recipient constellation. Similarly, as in \sprot{alg:hla_mux}, class $A$ indicates an optimal match, class $B$ a good match, and $Eq$ denotes that recipient and donor are in the same age group.

\emph{\mpc Cost.} \sprot{alg:age_mux} contains one comparison, one inversion, one \ANDGate{} gate, and three \MUXGate{} gates. As \lnsg{age:same} and \lnsg{age:ydor} are independent, similarly as the two \MUXGate{} gates in~\lnsg{age:mux}, the circuit depth is 3. Thus, this subprotocol is slightly more efficient in \bss than in \gc.

\begin{algorithm}
    \caption{$\cf{evalSex}(\seshce{s_d}{\bssform}$: int, $\seshce{s_r}{\bssform}$: int) $\rightarrow$ int}\label{alg:sex_mux}
    \begin{algorithmic}[1]
        \State $\seshce{eq}{\bssform} \gets \seshce{s_d}{\bssform} == \seshce{s_r}{\bssform} $ \label{sex:same}
        \State $\seshce{fdmr}{\bssform} \gets \seshce{s_d}{\bssform} \land \lnot \seshce{s_r}{\bssform}$ \label{sex:fdmr}
        \State\Return ${\seshce{fdmr}{\bssform}}$\textbf{ ? }\par
        \hskip\algorithmicindent${\left(\tern{\seshce{eq}{\bssform}}{\seshce{A}{\bssform}}{\seshce{0}{\bssform}}\right)}$\textbf{ : }${\left(\tern{\seshce{eq}{\bssform}}{\seshce{A}{\bssform}}{\seshce{B}{\bssform}}\right)}$ \label{sex:mux}
    \end{algorithmic}
\end{algorithm}

\noindent\textbf{Sex Comparison.} \sprot{alg:sex_mux} evaluates the compatibility of a donor and recipient based on their sex. It takes two secret shares $\langle s_d\rangle^B$ and $\langle s_r\rangle^B$ as input which represent the sex of the donor and recipient, respectively. In~\lnsg{sex:same}, the subprotocol determines if the pair shares the same sex. \lnsg{sex:fdmr} checks whether the donor is female and the recipient male. As final step, the output weight of this donor and recipient constellation is computed, i.e., the optimal combination ("Class A") with equal sex receives the highest weight, while a female donor and a male recipient are assigned the lowest weight (0).

\emph{\mpc Cost.} \sprot{alg:sex_mux} evaluates one comparison, one inversion, one \ANDGate, and three \MUXGate{} gates. As \lnsg{sex:same} and \lnsg{sex:fdmr} as well as two of the \MUXGate{} gates in \lnsg{sex:mux} are independent, we have a circuit depth of 3.
Thus, \sprot{alg:sex_mux} is slightly more efficient in \bss than in \gc.

\begin{algorithm}
    \caption{$\cf{evalWeight}(\seshce{w_d}{\bssform}$: int, $\seshce{w_r}{\bssform} $: int) $\rightarrow$ int}\label{alg:size_mux}
    \begin{algorithmic}[1]
        \State\Return $\tern{\seshce{w_d}{\bssform} < \seshce{w_r}{\bssform}}{\sesh{0}{\bssform}}{\seshce{A}{\bssform}}$
    \end{algorithmic}
\end{algorithm}

\noindent\textbf{Weight Comparison.} \sprot{alg:size_mux} evaluates the compatibility of a donor and recipient based on their weight. It takes two secret shares as input: $\langle w_d\rangle^B$ and $\langle w_r\rangle^B$ which represent the weight of the donor and recipient, respectively. If the donor weighs less than the recipient, it returns a secret shared $0$, otherwise, it indicates a good fit (i.e., class "A" w.r.t. criteria weight).

\emph{\mpc Cost.} We evaluate only one comparison gate. As the evaluation of a single comparison is more efficient in \gc than in \bss~\cite{dsz15}, \gc would be more efficient. However, the conversion cost outweigh this benefit which is why \bss is used for this subprotocol as in the previous comparison protocols.

\subsubsection*{Additional Protocols for the Cycle Computation}
\begin{algorithm}
    \caption{$\cf{removeWeights}(\seshce{compG}{\bssform}$: matrix) $\rightarrow$ matrix}\label{alg:unweighted_mux}
    \begin{algorithmic}[1]
        \State $\seshce{uG}{\assform} \gets \text{matrix} \in {\sesh{0}{\assform}}^{|\ce{pairs}|}$
        \For{$i = 0, \ldots, |\ce{pairs}| - 1$}
            \For{$j = 0, \ldots, |\ce{pairs}| - 1$}
                \State $\seshce{uG}{\assform}[i][j] \gets$\par
                \hskip\algorithmicindent$ \btoa(\tern{\seshce{compG}{\bssform}[i][j] > \sesh{0}{\bssform}}{\sesh{1}{\bssform}}{\sesh{0}{\bssform}}$ \label{uw:mux}
            \EndFor
        \EndFor
        \State\Return $\seshce{uG}{\assform}$
    \end{algorithmic}
\end{algorithm}

\noindent \textbf{Weight Removal.} In \sprot{alg:unweighted_mux}, we compute the unweighted compatibility graph which is used for determining the number of cycles for the desired cycle length. It takes the weighted compatibility graph \emph{compG} as input. The number of donor-recipient pairs $|\ce{pairs}|$ is public. In~\lnsg{uw:mux}, we remove the edge weights: If it is greater than~0, it is set to~$1$, otherwise to~$0$. As preparation for later processing, a conversion to \ass is done.

\emph{\mpc Cost.} \sprot{alg:unweighted_mux} evaluates $|\ce{pairs}|^2$ comparisons, \MUXGate{} gates, and conversions. The comparisons and \MUXGate{} gates are independent, which results in a circuit depth of 2. Due to the total number of \ANDGate{} gates, which is $2 \times |\ce{pairs}|$, this subprotocol is most efficient in \bss.

\begin{algorithm}
    \caption{$\cf{kNNSort}(\seshce{cyclesSet}{\gcform}$: vector of tuples, $k$: int) $\rightarrow$ vector of cycles} \label{alg:kNNSort_mux}
    \begin{algorithmic}[1]
        \State $\seshce{sortedW}{\gcform} \gets \emptyset$ \label{knn:init_start}
        \State $\seshce{sortedC}{\gcform} \gets \emptyset$
        \For{$i=0, \ldots, k$}
            \State $\seshce{sortedW}{\gcform}.\cf{append}(\sesh{0}{\gcform})$
            \State $\seshce{vertices}{\gcform} \gets \emptyset$
            \For{$j=0, \ldots, \ce{cLen}-1$}
                \State $\seshce{vertices}{\gcform}.\cf{append}(|\seshce{pairs}{\gcform}|)$
            \EndFor
            \State $\seshce{sortedC}{\gcform}.\cf{append}(\seshce{vertices}{\gcform})$
        \EndFor \label{knn:init_end}
        
        \For{$i=0,\ldots, |\ce{cyclesSet}|-1$} \label{knn:start_sort}
            \State $\seshce{sortedW}{\gcform}[k] \gets \seshce{cyclesSet}{\gcform}[i][0]$ 
            \State $\seshce{sortedC}{\gcform}[k] \gets \seshce{cyclesSet}{\gcform}[i][1]$ 
            \For{$j=0, \ldots, k-1$}
                \State $\seshce{sel}{\gcform} \gets \seshce{sortedW}{\gcform}[j] > \seshce{sortedW}{\gcform}[j-1]$ 
                \State $\seshce{tmp1}{\gcform} \gets \seshce{sortedW}{\gcform}[j]$
                \State $\seshce{tmp2}{\gcform} \gets \seshce{sortedW}{\gcform}[j-1]$
                \State $\seshce{sortedW}{\gcform}[j] \gets \tern{\seshce{sel}{\gcform}}{\seshce{tmp2}{\gcform}}{\seshce{tmp1}{\gcform}}$
                \State $\seshce{sortedW}{\gcform}[j-1] \gets \tern{\seshce{sel}{\gcform}}{\seshce{tmp1}{\gcform}}{\seshce{tmp2}{\gcform}}$
                \For{$l=0, \ldots, \ce{cLen}-1$}
                    \State $\seshce{tmp1}{\gcform} \gets \seshce{sortedC}{\gcform}[j][l]$
                    \State $\seshce{tmp2}{\gcform} \gets \seshce{sortedC}{\gcform}[j-1][l]$
                    \State $\seshce{sortedC}{\gcform}[j][l] \gets \tern{\seshce{sel}{\gcform}}{\seshce{tmp2}{\gcform}}{\seshce{tmp1}{\gcform}}$
                    \State $\seshce{sortedC}{\gcform}[j-1][l] \gets$\par
                    \hskip\algorithmicindent$ \tern{\seshce{sel}{\gcform}}{\seshce{tmp1}{\gcform}}{\seshce{tmp2}{\gcform}}$
                \EndFor
            \EndFor
        \EndFor \label{knn:end_sort}
        
        \State $\seshce{result}{\gcform} \gets \emptyset$
        \For{$i=0, \ldots,|\ce{cycles}|-1$}
            \State $\seshce{result}{\gcform}.\cf{append}(\text{tuple}(\seshce{sortedW}{\gcform}[i], \seshce{sortedC}{\gcform}[i]))$
        \EndFor
        \State\Return $\seshce{result}{\gcform}$
    \end{algorithmic}
\end{algorithm}

\noindent\textbf{kNN Sort Protocol.} Our next \mpc subprotocol shown in \sprot{alg:kNNSort_mux} is a $k$-nearest neighbor sort (a slightly adapted version of the protocol in~\cite{jarvinen19}) that identifies the $k$ most robust cycles (i.e., with the highest likelihood to result in successful transplantations).

It takes a secret shared vector of tuples $\ce{cyclesSet}$ with exchange cycles and their respective weights and $k$ as input. The length of cycles $\ce{cLen}$ is a public parameter.
First, the subprotocol iterates over all cycles in $|\ce{cyclesSet}|$ to perform an insertion sort. Each cycle and the respective weight are added to $\ce{sortedC}$ and $\ce{sortedW}$ if its weight is one of the $k$ highest weights (cf.~\lnpl{knn:start_sort} to~\ref{knn:end_sort}). Thus, the final $\ce{sortedW}$ and $\ce{sortedC}$ are sorted in decreasing order with respect to the weights of cycles.

\noindent\emph{\mpc Cost.} This subprotocol evaluates $|\ce{cyclesSet}| \times k$ comparisons and $|\ce{cyclesSet}| \times k \times (1+\ce{cLen})$ \MUXGate{} gates. It is most efficient in \gc due to depth of the circuit determined by the number of \ANDGate{} gates.

\noindent \textbf{Duplicate Removal.} \sprot{alg:removeDuplicates_mux} removes all duplicated exchange cycles and outputs the remaining $|\ce{unique}|=\lfloor \frac{|\ce{cycles}|}{\ce{cLen}} \rfloor$ cycles.

It takes a secret shared vector of tuples $\ce{sortedCycles}$ as input, which contains cycles and weights sorted according to the respective weights (i.e., the output by \sprot{alg:kNNSort_mux}). The number of existing cycles $|\ce{cycles}|$, the number of unique cycles $|\ce{unique}|$, and the cycle length $\ce{cLen}$ are public~parameters. For each cycle $c1$ in $\ce{sortedCycles}$, it is checked if it is equal to any other cycle $c2$ (cf.~\lnpl{rD:CheckDuplicate_begin} to~\ref{rD:CheckDuplicate_end}). If this is the case, its weight is set to $0$ (cf.~Line~\ref{rD:select_weight}).
To each equality, it is evaluated if the vertex of $c1$ at index $l$ and the vertex of $c2$ at index $(l + k) \mod\ce{cLen}$ are identical
(cf.~Line~\ref{rD:shift}).  
With \sprot{alg:kNNSort_mux}, $\ce{sortedCycles}$ is sorted and only the $|\ce{unique}|$ cycles with the highest weight are returned. The number of unique cycles is $|\ce{unique}| = \lfloor \frac{|\ce{cycles}|}{\ce{cLen}} \rfloor$.

\noindent\emph{\mpc Cost.} \sprot{alg:removeDuplicates_mux} has $|\ce{cycles}| \times \sum_{i = 0}^{|\ce{cycles}|}\left(\ce{cLen} \times (\ce{cLen}-1)\right)$ comparisons and \ANDGate{} gates, $|\ce{cycles}| \times \sum_{i = 0}^{|\ce{cycles}|}\left(\ce{cLen}-1\right)$ \ORGate{} gates, $|\ce{cycles}|$ \MUXGate{} gates. Including \sprot{alg:kNNSort_mux}, this results in $|\ce{cycles}| \times |\ce{unique}|$ comparison and \MUXGate{} gates, and an additional $|\ce{cycles}| \times |\ce{unique}| \times (1 +\ce{cLen})$ \MUXGate{} gates. This subprotocol is most efficient in \gc due to the depth of the circuit created by \ANDGate{} gates.

\begin{algorithm}
    \caption{$\cf{removeDuplicates}(\seshce{sortedCycles}{\gcform}$: vector of tuples) $\rightarrow$ vector of cycles} \label{alg:removeDuplicates_mux}
    \begin{algorithmic}[1]
        \For{$i=0, \ldots,|\ce{cycles}|-1$} \label{rD:foreach}
            \State $\seshce{c1}{\gcform} \gets \seshce{sortedCycles}{\gcform}[i][1]$
            \State $\seshce{combDup}{\gcform} \gets \sesh{0}{\gcform}$
             \For{$j=0: i$}
                \State $\seshce{c2}{\gcform} \gets \seshce{sortedCycles}{\gcform}[j][1]$
                \For{$k=1, \ldots, \ce{cLen}-1$}  \label{rD:CheckDuplicate_begin}
                    \State $\seshce{duplicate}{\gcform} \gets \sesh{1}{\gcform}$
                    \For{$l=0, \ldots, \ce{cLen}-1$}
                        \State $\seshce{same}{\gcform} \gets $\par
            \hskip\algorithmicindent$\seshce{c1}{\gcform}[l] == \seshce{c2}{\gcform}[(l+k) \mod \ce{cLen}]$ \label{rD:shift}
                        \State $\seshce{duplicate}{\gcform} \gets \seshce{duplicate}{\gcform} \land \seshce{same}{\gcform}$
                    \EndFor
                    \State $\seshce{combDup}{\gcform} \gets \seshce{combDup}{\gcform} \lor \seshce{duplicate}{\gcform}$
                \EndFor \label{rD:CheckDuplicate_end}
            \EndFor
           \State $\seshce{sortedCycles}{\gcform}[i][0] \gets $\par
            \hskip\algorithmicindent${\seshce{isDuplicate}{\gcform}}\text{\textbf{ ? }}{\sesh{0}{\gcform}}\text{\textbf{ : }}{\seshce{sortedCycles}{\gcform}[i][0]}$  \label{rD:select_weight}
        \EndFor\label{rD:foreachend}
        \State\Return $\cf{kNNSort}(\seshce{sortedCycles}{\gcform}, |\ce{unique}|)$ \label{rD:sort}
    \end{algorithmic}
\end{algorithm}

\begin{algorithm}[H]
    \caption{$\cf{\#TotalCycles}$() $\rightarrow$ int} \label{alg:total_cycles}
    \begin{algorithmic}[1]
        \State $ |\ce{allCycles}| \gets |\ce{pairs}|$
        \For{$i=1, \ldots, \ce{cLen} -1$} 
            \State $|\ce{allCycles}| \gets |\ce{allCycles}| \cdot (|\ce{pairs}| - i)$
        \EndFor
        \State\Return $|\ce{allCycles}|$
    \end{algorithmic}
\end{algorithm}

\noindent\textbf{Total Number of Cycles.} \sprot{alg:total_cycles} computes the maximum number of cycles that can exist in the compatibility graph. Each vertex must appear at most once in a cycle which limits the number of possible cycles. As the numbers of pairs $|\ce{pairs}|$ and the cycle length $\ce{cLen}$ are public, computation can be done on plaintext. 

\subsubsection*{Additional Protocols for the Solution Evaluation}
\begin{algorithm}
    \caption{$\cf{disjointSet}(\seshce{cycles}{\bssform}$: vector of tuples, $\seshce{cCycle}{\bssform}$: vector, $count$: int) $\rightarrow$ Boolean} \label{alg:disjoint}
    \begin{algorithmic}[1]
       \State $\seshce{disJ}{\bssform} \gets \emptyset$
       \For{$i=0, \ldots, \ce{count}-1 $}
            \State $\seshce{c}{\bssform} \gets \seshce{cycles}{\bssform}[i][1]$
           \For{$j = 0, \ldots, \ce{cLen}-1$}
                \For{$k = 0, \ldots, \ce{cLen}-1$} \label{dj:tree_start}
                   \State $\seshce{tmp}{\bssform} \gets \seshce{c}{\bssform}[j] == \seshce{cCycle}{\bssform}[k]$\label{prot:disj1}
                   \State $\seshce{disJ}{\bssform}.\cf{append}(\seshce{tmp}{\bssform})$ 
                \EndFor \label{dj:tree_end}
            \EndFor
            \State $\seshce{disJ}{\bssform} \gets \cf{ORTREE}(\seshce{disJ}{\bssform})$ \label{prot:disj2}
        \EndFor
        \State\Return $\lnot \seshce{disJ}{\bssform}[0]$\label{prot:disj_return}
    \end{algorithmic}
\end{algorithm}

\noindent \textbf{Disjoint Cycles.} \sprot{alg:disjoint} computes whether a cycle \texttt{cCycle} does not join vertices with other cycles of a set of cycles \texttt{cycles}. It takes as input the set of secret shared cycles \texttt{cycles}, the secret shared cycle \texttt{cCycle}, and the number of cycles in \texttt{cycles} count. If another cycle shares a vertex with \texttt{cCycle}, \texttt{disJ} is set to $1$ (cf.~Line~\ref{prot:disj2}). In \lnsg{prot:disj_return}, we invert the result for further evaluation.

\noindent\emph{\mpc Cost.} In this subprotocol, we evaluate $|\ce{cycles}| \times \ce{cLen}$~(cf.~Line~\ref{prot:disj1}). In Line~\ref{prot:disj2}, we evaluate $\log_2(\ce{cLen})$ \ORGate{} gates. At the end, we evaluate one \XORGate{} gate. As the number of total \ANDGate{} gates is greater than the depth of the circuit, this subprotocol is most efficient in \bss.

\begin{algorithm}[H]
    \caption{$\cf{findMaximumSet}(\seshce{cyclesSets}{\gcform}$: vector of tuples, $\seshce{cycleW}{\gcform}$: vector) $\rightarrow$ tuple} \label{alg:max_set_mux}
    \begin{algorithmic}[1]
        \State $ \seshce{weights}{\gcform} \gets \emptyset$
        \State $ \seshce{tmp}{\gcform} \gets \emptyset$
        \For{$i=0, 1$}
            \State $\seshce{weights}{\gcform}.\cf{append}(\sesh{0}{\gcform})$
            \State $\seshce{sets}{\gcform} \gets \emptyset$
            \For{$j=0, \ldots |\ce{unique}|-1$}
                \State $\seshce{vertices}{\gcform} \gets \emptyset$
                \For{$j=0, \ldots \ce{cLen}-1$}
                    \State $\seshce{vertices}{\gcform}.\cf{append}(\seshce{|\ce{pairs}|}{\gcform})$
                \EndFor
                \State $\seshce{tmp}{\gcform}.\cf{append}(\seshce{vertices}{\gcform})$
            \EndFor
            \State $\seshce{sets}{\gcform}.\cf{append}(\seshce{tmp}{\gcform})$
        \EndFor
        
        \For{$i=0,\ldots,|\ce{unique}|-1$}
            \State $\seshce{weights}{\gcform}[1] \gets \seshce{cycleW}{\gcform}[i]$
            \State $\seshce{sets}{\gcform}[1] \gets \seshce{cycleSets}{\gcform}[i]$
            \State $\seshce{sel}{\gcform} \gets \seshce{weights}{\gcform}[1] > \seshce{weights}{\gcform}[0]$
            
            \State $\seshce{tmp1}{\gcform} \gets \seshce{weights}{\gcform}[1]$
            \State $\seshce{tmp2}{\gcform} \gets \seshce{weights}{\gcform}[0]$
            \State $\seshce{weights}{\gcform}[1] \gets \tern{\seshce{sel}{\gcform}}{\seshce{tmp2}{\gcform}}{\seshce{tmp1}{\gcform}}$
            \State $\seshce{weights}{\gcform}[0] \gets \tern{\seshce{sel}{\gcform}}{\seshce{tmp1}{\gcform}}{\seshce{tmp2}{\gcform}}$
           
                \For{$j=0, \ldots, |\ce{unique}|-1$}
                    \State $\seshce{tmp1}{\gcform} \gets \seshce{sets}{\gcform}[1][j]$
                    \State $\seshce{tmp2}{\gcform} \gets \seshce{sets}{\gcform}[0][j]$
                    \State $\seshce{sets}{\gcform}[1][j] \gets \tern{\seshce{sel}{\gcform}}{\seshce{tmp2}{\gcform}}{\seshce{tmp1}{\gcform}}$
                    \State $\seshce{sets}{\gcform}[0][j] \gets \tern{\seshce{sel}{\gcform}}{\seshce{tmp1}{\gcform}}{\seshce{tmp2}{\gcform}}$
                \EndFor
        \EndFor
        
        \State\Return$(\seshce{weights}{\gcform}[0], \seshce{sets}{\gcform}[0])$
    \end{algorithmic}
\end{algorithm}

\textbf{Maximum Set.} \sprot{alg:max_set_mux} computes the set of cycles with the highest sum of weights, thus, the set of cycles with the highest probability for successful transplantations. Note that we do not compute the globally optimal solution but a local optimum.

The subprotocol takes a secret shared vector of tuples $\ce{cycles}$ and a secret shared vector $\ce{weights}$ as input. $\ce{cycles}$ contains all sets of disjoint cycles and $\ce{weights}$ contains the respective weights of the each set. The number of pairs $|\ce{pairs}|$ and the number of unique cycles $|\ce{unique}|$ are public parameters. This subprotocol is a slight variation of~\sprot{alg:kNNSort_mux} adapted to here used data structures. The parameter $k$ is fixed to $1$ since we look for the set with the highest weight.

\noindent\emph{\mpc Cost.} This protocols evaluates $|\ce{unique}|$ comparison and $2|\ce{unique}|^2 + 2|\ce{unique}|$ \MUXGate{} gates. Due to the  large number of \MUXGate{} gates in combination with the number of \ANDGate{} gates determining the depth of the circuit, it is most efficient in \gc sharing. 

\subsection*{Communication Improvement with ABY2.0}\label{appendix:aby2}

Implementing \ourname{} in ABY2.0~\cite{patra2021aby2} can further decrease the communication cost. As the respective protocols were implemented only very recently in MOTION2NX~\cite{BCS21PriMLNeurIPS}, we use ABY~\cite{dsz15} in our benchmarks to show practicality and additionally discuss the improvements that can be achieved with ABY2.0 in the~following. 

ABY2.0's improvements for secure multiplication with two inputs decreases the communication of the first and second part, the compatibility matching and the cycle computation.
The improvements to conversions between different sharing types additionally benefit the first and the second part as these parts contain the most conversions between \bss{} and \ass. Further, the optimizations for matrix multiplications are beneficial for the second part.

Concretely, the communication of Protocol~\ref{alg:comapt_graph_mux} decreases from $3 \times \ell^2 + 24 \times \ell + 2 \times \ell \times \kappa$ bits to $23 \times \ell + \ell \times \kappa$ bits in every iteration of the inner loop (without considering the subprotocols). Similarly, Protocol~\ref{alg:det_cycles_general}'s communication can be reduced from $\ell \times (\frac{\ell}{2} + 2 \times \kappa + 4 \times |\ce{pairs}|^3 + 1.5)$ bits to $\ell \times (\kappa + 2 \times |\ce{pairs}|^3 + 3)$ bits, where $\ell$ is the bitlength and $\kappa$ is the security parameter.

\subsection*{ABY Security Assumptions and Guarantees}
The ABY \mpc framework~\cite{dsz15} provides mixed-abstraction building blocks
for the creation of highly efficient hybrid-protocol \mpc applications in a
semi-honest adversary setting. Independent of the specific circuit design, a
number of base-OTs are executed in the beginning to setup OT Extensions. The
used base-OT primitive~\cite{naorEfficientOblivious2001} guarantees security
under the Computational Diffie-Hellman (CDH) hardness assumption. Being closely
related to the discrete logarithm problem, this assumption is known to not be
quantum resistant. The OT Extension~\cite{ishaiExtendingObliviousTransfers2003,
asharovMoreEfficient2017} primitive uses fixed-key AES modeled as a random permutation.
While still considered secure in a semi-honest setting, theoretical
attacks in the active security setting have been
demonstrated~\cite{guoEfficientSecure2020}. Furthermore, ABY relies on the
random oracle assumption, implemented by the SHA256 hash function. Similarly,
Yao's Garbled Circuits~\cite{yaoHowGenerate1986} (denoted by \gc in our
protocols) directly rely on the random permutation assumptions, while Arithmetic
and Boolean Sharing~\cite{goldreichHowPlay1987} (denoted by \ass/\bss in our
protocols) indirectly rely on those assumptions as a source of randomness. Those
protocols can achieve information-theoretical security given a true correlated
randomness source.

\subsection*{Detailed Benchmark Results}
\tabs{tab:comparison-cl2-p1}~to~\ref{tab:comparison-cl2-p3} show
the detailed benchmark results for the setup and online phase in all three
described network settings (A:\@ LAN + \Gbits{10}, B:\@ LAN + \Gbits{1}, C: WAN)
and a cycle length of $L=2$.
\tabs{tab:comparison-cl3-p1}~and~\ref{tab:comparison-cl3-p2} show the results for a cycle
length of $L=3$.

\tab{tab:comparison-matching}, finally, compares the benchmark results of both reduced medical compatibility factor set and the full set. This benchmark was performed in the two network settings A and C, as above.

\pgfplotstableset{
  col sep=comma,
  columns/pairs/.style={
    int detect,
    column type=r,
    column name={Pairs}
  },
  columns/setupComm/.style={
    fixed,
    precision=1,
    column type = r,
    divide by={1048576}, 
    column name={Setup},
  },
  columns/onlineComm/.style={
    fixed,
    precision=1,
    column type = r,
    divide by={1048576}, 
    column name={Online},
  },
  empty cells with={--}, 
  every last row/.style={after row=\bottomrule}
}

\pgfplotsinvokeforeach{%
  setup.mean.lan10,online.mean.lan10,%
  setup.mean.lan1,online.mean.lan1,%
  setup.mean.wan,online.mean.wan%
} {%
  \pgfplotstableset{
    columns/#1/.style={
      fixed relative,
      precision=2,
      dec sep align,
      divide by={1000}
    }
  }
}

\pgfplotstableset{columns/setup.mean.lan10/.append style={column name={A}}}
\pgfplotstableset{columns/online.mean.lan10/.append style={column name={A}}}
\pgfplotstableset{columns/setup.mean.lan1/.append style={column name={B}}}
\pgfplotstableset{columns/online.mean.lan1/.append style={column name={B}}}
\pgfplotstableset{columns/setup.mean.wan/.append style={column name={C}}}
\pgfplotstableset{columns/online.mean.wan/.append style={column name={C}}}

\newcommand{\AddComparisonTable}[2][]{%
  \pgfplotstabletypeset[columns={
    pairs,setupComm,onlineComm,%
    setup.mean.lan10,setup.mean.lan1,setup.mean.wan,%
    online.mean.lan10,online.mean.lan1,online.mean.wan%
  },
  every head row/.style={
    before row = {
      \toprule
      Pairs & \multicolumn{2}{c}{Comm.\ [MiB]} & \multicolumn{6}{c}{Setup Phase [s]} & \multicolumn{6}{c}{Online Phase [s]} \\
      \cmidrule(lr){2-3} \cmidrule(lr){4-9} \cmidrule(lr){10-15}
    },
    after row=\midrule
  },
  #1
  ]{#2}
}

\newcommand{\AddMatchingTable}[2][]{%
  \pgfplotstabletypeset[columns={
    pairs,setupComm,onlineComm,%
    setup.mean.lan10,setup.mean.wan,%
    online.mean.lan10,online.mean.wan%
  },
  every head row/.style={
    before row = {
      \toprule
      Pairs & \multicolumn{2}{c}{Comm.\ [MiB]} & \multicolumn{4}{c}{Setup Phase [s]} & \multicolumn{4}{c}{Online Phase [s]} \\
      \cmidrule(lr){2-3} \cmidrule(lr){4-7} \cmidrule(l){8-11}
    },
    after row=\midrule
  },
  #1
  ]{#2}
}

\begin{table*}[h!tbp]
  \centering
  \setlength{\tabcolsep}{4.5pt}
  \caption{Comparison of the communication costs and setup and online runtimes of \ourname{} for the three networking
  configurations A:\@ LAN + \Gbits{10}, B:\@ LAN + \Gbits{1}, C:\@ WAN, and for cycle
length $L=2$. This table contains the aggregated total costs and the individual costs of Phase
1 (Compatibility Matching).}%
\label{tab:comparison-cl2-p1}

\pgfplotstablevertcat{\allCompTable}{comparison_cl2_total.csv}
\pgfplotstablevertcat{\allCompTable}{comparison_cl2_p1.csv}

\AddComparisonTable[%
  every row no 0/.append style={before row={%
      \addlinespace
      \multicolumn{7}{l}{\emph{Total}}&&& \\
      \midrule
  }},
  every row no 24/.append style={before row={%
      \addlinespace
      \multicolumn{7}{l}{\emph{Phase 1: Compatibility Matching}}&&& \\
      \midrule
  }}%
]{\allCompTable}
\end{table*}
\begin{table*}[h!tbp]
  \centering
  \setlength{\tabcolsep}{4.5pt}
  \caption{Comparison of the communication costs and setup and online runtimes of \ourname{} for the three networking
  configurations A:\@ LAN + \Gbits{10}, B:\@ LAN + \Gbits{1}, C:\@ WAN, and for cycle
length $L=2$. This table contains individual costs of Phase 2 and 3 (Cycle Computation and Evaluation).}%
\label{tab:comparison-cl2-p2}

\pgfplotstablevertcat{\allCompTable}{comparison_cl2_p2.csv}
\pgfplotstablevertcat{\allCompTable}{comparison_cl2_p3.csv}

\AddComparisonTable[%
  every row no 0/.append style={before row={%
      \addlinespace
      \multicolumn{7}{l}{\emph{Phase 2: Cycle Computation}}&&& \\
      \midrule
  }},
  every row no 24/.append style={before row={%
      \addlinespace
      \multicolumn{7}{l}{\emph{Phase 3: Cycle Evaluation}}&&& \\
      \midrule
  }}%
]{\allCompTable}
\end{table*}
\begin{table*}[h!tbp]
  \centering
  \setlength{\tabcolsep}{4.5pt}
  \caption{Comparison of the communication costs and setup and online runtimes of \ourname{} for the three networking
  configurations A:\@ LAN + \Gbits{10}, B:\@ LAN + \Gbits{1}, C:\@ WAN and for cycle
length $L=2$. This table contains individual costs of Phase 4 (Solution Evaluation).}%
\label{tab:comparison-cl2-p3}

\pgfplotstablevertcat{\allCompTable}{comparison_cl2_p4.csv}

\AddComparisonTable[%
  every row no 0/.append style={before row={%
      \addlinespace
      \multicolumn{7}{l}{\emph{Part 4: Solution Evaluation}}&&& \\
      \midrule
  }}%
]{\allCompTable}
\end{table*}
\clearpage
\begin{table*}[h!tbp]
  \centering
  \setlength{\tabcolsep}{4.5pt}
  \caption{Comparison of the communication costs and setup and online runtimes of \ourname{} for the three networking
  configurations A:\@ LAN + \Gbits{10}, B:\@ LAN + \Gbits{1}, C:\@ WAN and for cycle
length $L=3$. This table contains the aggregated total costs and the individual costs of Phases
1 and 2 (Compatibility Matching and Cycle Computation).}%
\label{tab:comparison-cl3-p1}

\pgfplotstablevertcat{\allCompTable}{comparison_cl3_total.csv}
\pgfplotstablevertcat{\allCompTable}{comparison_cl3_p1.csv}
\pgfplotstablevertcat{\allCompTable}{comparison_cl3_p2.csv}

\AddComparisonTable[%
  every row no 0/.append style={before row={%
      \addlinespace
      \multicolumn{7}{l}{\emph{Total}}&&& \\
      \midrule
  }},
  every row no 16/.append style={before row={%
      \addlinespace
      \multicolumn{7}{l}{\emph{Phase 1: Compatibility Matching}}&&& \\
      \midrule
  }},
  every row no 32/.append style={before row={%
      \addlinespace
      \multicolumn{7}{l}{\emph{Phase 2: Cycle Computation}}&&& \\
      \midrule
  }}%
]{\allCompTable}
\end{table*}
\begin{table*}[h!tbp]
  \centering
  \setlength{\tabcolsep}{4.5pt}
  \caption{Comparison of the communication costs and setup and online runtimes of \ourname{} for the three networking
  configurations A:\@ LAN + \Gbits{10}, B:\@ LAN + \Gbits{1}, C:\@ WAN and for cycle
length $L=3$. This table contains the individual costs of Phases 3 and 4 (Cycle and Solution Evaluation).}%
\label{tab:comparison-cl3-p2}

\pgfplotstablevertcat{\allCompTable}{comparison_cl3_p3.csv}
\pgfplotstablevertcat{\allCompTable}{comparison_cl3_p4.csv}

\AddComparisonTable[%
  every row no 0/.append style={before row={%
      \addlinespace
      \multicolumn{7}{l}{\emph{Phase 3: Cycle Evaluation}}&&& \\
      \midrule
  }},
  every row no 16/.append style={before row={%
      \addlinespace
      \multicolumn{7}{l}{\emph{Phase 4: Solution Evaluation}}&&& \\
      \midrule
  }}%
]{\allCompTable}
\end{table*}

\begin{table*}[h!tbp]
  \centering
  \setlength{\tabcolsep}{4.5pt}
  \caption{Comparison of the setup and online runtimes of \ourname{} for the reduced medical factor compatibility matching and the full set in the two main networking configurations A:\@ LAN + \Gbits{10}, C:\@ WAN.}%
\label{tab:comparison-matching}

\pgfplotstablevertcat{\allCompTable}{comparison_matching_reduced.csv}
\pgfplotstablevertcat{\allCompTable}{comparison_matching_all.csv}

\AddMatchingTable[%
  every row no 0/.append style={before row={%
      \addlinespace
      \multicolumn{7}{l}{\emph{Reduced Medical Factor Set}}&&& \\
      \midrule
  }},
  every row no 14/.append style={before row={%
      \addlinespace
      \multicolumn{7}{l}{\emph{Full Medical Factor Set}}&&& \\
      \midrule
  }}%
]{\allCompTable}
\end{table*}

\end{backmatter}
\end{document}